\documentclass[useAMS,usenatbib,fleqn]{mn2e}
\usepackage[pdftex]{graphicx}
\usepackage{amsmath,amssymb,hyperref}
\usepackage{times}
\usepackage[T1]{fontenc}

\title[Polarization leakage in Epoch of Reionization windows]{Polarization leakage in Epoch of Reionization windows:\\ 
I. LOFAR  observations of the 3C196 field}
\author[K. M. B. Asad et al.]
{K. M. B. Asad$^1$\thanks{E-mail: khan@astro.rug.nl},
L. V. E. Koopmans$^{1}$,
V. Jeli\'{c}$^{1,2,3}$,
V. N. Pandey$^2$,
A. Ghosh$^1$,
\newauthor
F. B. Abdalla$^{4,5}$,
G. Bernardi$^6$,
M. A. Brentjens$^2$,
A. G. de Bruyn$^{1,2}$,
S. Bus$^1$,
\newauthor
B. Ciardi$^7$,
E. Chapman$^4$,
S. Daiboo$^8$,
E. R. Fernandez$^1$,
G. Harker$^4$,
I. T. Iliev$^9$,
\newauthor
H. Jensen$^{10}$,
O. Martinez-Rubi$^1$,
G. Mellema$^{10}$,
M. Mevius$^{1,2}$,
A. R. Offringa$^{1,2}$,
\newauthor
A. H. Patil$^1$,
J. Schaye$^{11}$,
R. M. Thomas$^1$,
S. van der Tol$^{2,11}$,
H. K. Vedantham$^1$,
\newauthor
S. Yatawatta$^{1,2}$
and S. Zaroubi$^1$\\ \\
$^1$Kapteyn Astronomical Institute, University of Groningen, PO Box 800, NL-9700 AV Groningen, the Netherlands\\
$^2$ASTRON, PO Box 2, NL-7990 AA Dwingeloo, the Netherlands\\
$^3$Ru{\dj}er Bo\v{s}kovi\'{c} Institute, Bijeni\v{c}ka cesta 54, 10000 Zagreb, Croatia\\
$^4$Department of Physics \& Astronomy, University College London, Gower Street, London WC1E 6BT, UK\\
$^5$Department of Physics and Electronics, Rhodes University, PO Box 94, Grahamstown, 6140, South Africa\\
$^6$SKA SA, 3rd Floor, The Park, Park Road, Pinelands 7405, South Africa\\
$^7$Max-Planck Institute for Astrophysics, Karl-Schwarzschild-Strasse 1, D-85748 Garching bei M\"{u}nchen, Germany\\
$^8$Observatoire de Paris, 61 avenue de l'Observatoire Paris, 75014 France\\
$^9$Astronomy Centre, Department of Physics \& Astronomy, Peven
sey II Building, University of Sussex, Falmer, Brighton BN1
9QH, UK\\
$^{10}$Department of Astronomy and Oskar Klein Centre, Stockholm University, AlbaNova, SE-10691 Stockholm, Sweden\\
$^{11}$Leiden Observatory, Leiden University, PO Box 9513, NL-2300 RA Leiden, the Netherlands\\
}

\begin{document}
\label{firstpage}
\date{Accepted . Received ; in original form }

\pagerange{\pageref{firstpage}--\pageref{lastpage}} \pubyear{2014}

\maketitle

\begin{abstract}
Detection of the 21-cm signal coming from the epoch of reionization (EoR) is challenging especially because, even after removing the foregrounds, the residual Stokes $I$ maps contain leakage from polarized emission that can mimic the signal.
Here, we discuss the instrumental polarization of LOFAR and present realistic simulations of the leakages between Stokes parameters.
From the LOFAR observations of polarized emission in the 3C196 field, we have quantified the level of polarization leakage caused by the nominal model beam of LOFAR, and compared it with the EoR signal using power spectrum analysis.
We found that at 134--166 MHz, within the central 4$^\circ$ of the field the $(Q,U)\rightarrow I$ leakage power is lower than the EoR signal at $k<0.3$ Mpc$^{-1}$.
The leakage was found to be localized around a Faraday depth of 0, and the rms of the leakage as a fraction of the rms of the polarized emission was shown to vary between 0.2--0.3\%, both of which could be utilized in the removal of leakage.
Moreover, we could define an `EoR window' in terms of the polarization leakage in the cylindrical power spectrum above the PSF-induced wedge and below $k_\parallel\sim 0.5$ Mpc$^{-1}$, and the window extended up to $k_\parallel\sim 1$ Mpc$^{-1}$ at all $k_\perp$ when 70\% of the leakage had been removed.
These LOFAR results show that even a modest polarimetric calibration over a field of view of $\lesssim 4^\circ$ in the future arrays like SKA will ensure that the polarization leakage remains well below the expected EoR signal at the scales of 0.02--1 Mpc$^{-1}$.
\end{abstract}

\section{Introduction}
Five phases of the large-scale universe are imprinted on Hydrogen: (i) the primordial phase before redshift $z\sim 1100$---when the universe was a hot, dense plasma---that ended when protons recombined with electrons releasing the photons that we detect today as a $\sim 2.7$ K signal known as the cosmic microwave background (CMB); (ii) the `Dark Ages' ($1100\gtrsim z\gtrsim 30$) when the baryonic universe contained mostly neutral Hydrogen and freely moving photons; (iii) the `Cosmic Dawn' ($30\gtrsim z\gtrsim 12$) when the first structures formed; (iv) the `Epoch of Reionization' (EoR; $12\gtrsim z\gtrsim 6.5$) when high-energy photons emitted by the first sources reionized the Hydrogen in the intergalactic medium; and (v) the current phase ($z\lesssim 6.5$) when almost all Hydrogen in the universe are ionized \citep[e.g.][]{fu,me,za}.

The aforementioned highly uncertain boundaries of the EoR have been approximated using indirect probes, e.g. CMB polarization at the high-$z$ end \citep[e.g.][]{pa07} and absorption features in quasar spectra at the low-$z$ end \citep[e.g.][]{fa}.
However, the new generation low-frequency, wide-bandwidth radio interferometers have the potential to directly detect the 21-cm radiation emitted by neutral Hydrogen during the EoR, redshifted to the wavelengths of around 1.5--3 m (corresponding to 200--100 MHz), as a differential brightness with respect to the CMB.
There are several ongoing and planned experiments to detect the EoR signal using radio arrays:
Giant Metrewave Radio Telescope (GMRT)\footnote{http://gmrt.ncra.tifr.res.in/}, 
Low Frequency Array (LOFAR)\footnote{http://www.lofar.org/}, 
Murchison Widefield Array (MWA)\footnote{http://www.mwatelescope.org/}, 
Precision Array for Probing the EoR (PAPER)\footnote{http://eor.berkeley.edu/}, 
21-cm Array (21CMA)\footnote{http://21cma.bao.ac.cn}, and the planned 
Square Kilometre Array (SKA)\footnote{http://www.skatelescope.org/}.

In order to detect the EoR, the effect of all other signals, e.g. the Galactic and extragalactic foregrounds, has to be excluded from the observed data; spatial fluctuations of the Galactic foreground can be 2-3 orders of magnitude higher than that of the EoR signal \citep{be09,be10,po13} which is around 10 mK within the redshifts 6--10 at $3'$ resolution \citep{pa14}.
However, even after removing the foregrounds with high accuracy the system noise after even hundreds of hours of integration will be an order of magnitude higher than the signal, thereby forcing us to aim for a statistical detection of the signal.
One of the methods for detecting the EoR signal statistically entails removing the foregrounds with high accuracy and then measuring the power spectrum of the residual which depends heavily on a proper understanding of the systematic and the random (noise) errors associated with the observing instrument and foreground removal 
\citep[e.g.][]{di15,lia,lib,be13,ch13,ve13,mo12,pa12,be10,ha10,je08}.

In this paper we address the systematic errors due to polarized foregrounds associated with the EoR experiment being conducted using LOFAR (the LOFAR-EoR project).
After taking out the bright extragalactic foreground, i.e. the resolved point sources, the Galactic foreground can be removed utilizing the fact that the EoR signal has significant correlated structure along the frequency---or equivalently the redshift---axis while the Galactic diffuse foreground is spectrally smooth in Stokes $I$.
However, the Faraday rotated polarized Galactic foreground is not always smooth along frequency and hence a leakage of the polarized emission into Stokes $I$ might mimic the EoR signal \citep[e.g.][]{je10}.
Systematic errors can cause this leakage in two different ways: direction independent (DI) and direction dependent (DD).
Non-orthogonal or rotated feeds of an antenna of an interferometer can cause $Q$ to leak into $I$ and vice versa while cross-talk between two feeds can cause mixing between all 4 Stokes parameters.
As these are DI errors, they can be corrected with high accuracy using traditional self-calibration.
However, the DD errors (DDE) caused by the time-frequency-baseline dependent primary beams cannot be corrected so easily.
In the latter case, an ellipticity of the beam can cause $I\leftrightarrow Q$ mixing while cross-polarization between two orthogonal components of the beam can mix all Stokes parameters.

\citet{ca} calculated a full polarization Mueller matrix to account for the look-direction dependent polarization aberration inherent in a dipole interferometer due to the fact that a source sees different projections of a dipole at different times.
\citet{je10} used this Mueller matrix to calculate the amount of leakage to be expected over the field of view of LOFAR and found that the leakage should be 0.1-0.7\% at 138 MHz within a $5^\circ\times 5^\circ$ patch of sky around the zenith and should increase to 2-20\% for an elevation of $45^\circ$.
If the polarized intensity is $\sim 1$ K, then a 1.5\% leakage would give a polarized emission of $\sim 15$ mK in Stokes $I$ which is comparable to the EoR signal.
\citet{mo} simulated the sky with randomly generated Faraday rotated, polarized point sources and found that the power of $Q\rightarrow I$ leakage due to the model beam of PAPER that has a FWHM of around $45^\circ$ at 150 MHz is of the order of thousands of [mK]$^2$ which is several orders of magnitude higher than the expected EoR signal power.
Their result turned out to be pessimistic because of their choice of the model; in reality, point sources are much more weakly polarized at low frequencies \citep{be13}.

Here, we predict the level of polarization leakage to be expected in the 3C196 window of the LOFAR-EoR experiment using reasonable models of the field and the model beam of LOFAR produced by \citet{ha11} using an electromagnetic simulation of the ASTRON Antenna Group\footnote{M. J. Arts; http://www.astron.nl}, and also test some leakage-correction strategies.
The paper is organized as follows.
Section \ref{s:form} revisits the mathematical formalism of a radio interferometer and describes the DI errors and the LOFAR beam-related DD errors within the context of this formalism.
Formalisms used for calibration, imaging, flux conversion, RM synthesis and power spectrum analysis are also described briefly.
In section \ref{s:egal} we describe the pipeline and setup of the simulations of extragalactic point sources and present three different results: effect of DI errors and the accuracy of self-calibration in this case, effect of DD-errors and a possible DDE correction strategy, and finally errors due to self-calibration with incomplete sky models.
Pipeline, setup and results of the simulation of Galactic foreground are presented in section \ref{s:gal}, where we show the results of rotation measure synthesis and power spectrum analysis, compare the power spectra of the leakage and the expected EoR signal, and test a potential leakage removal method.
In section \ref{s:summ} we give a summary of the paper, discuss some of the assumptions and limitations briefly and, finally, list the major conclusions of this paper.

\section{Formalism}\label{s:form}
\subsection{Mathematical model of a radio interferometer}
Here, we give an outline of the mathematical model of a radio interferometer and refer the readers to \citet{ha96,sm11a} for a detail description.

Consider a quasi-monochromatic electromagnetic wave propagating through space from a single point source.
Using the Cartesian coordinate system $xyz$ where the signal propagates along $z$ direction, the signal, at a specific point in time ($t$) and space, can be described by the complex vector $\mathcal{E}(x,y,t)$ and transformations (e.g. contaminations) of this signal along its path can be represented by $2\times 2$ Jones matrices.
Assuming all such transformations to be linear, a cumulative Jones matrix ($\mathbf{J}$) can be constructed from the products of the matrices.
The signal detected by our telescope will be the intrinsic signal multiplied by this cumulative matrix, mathematically\footnote{In this paper vectors are represented by calligraphy, matrices by bold and scalars by normal typefaces.} $\mathcal{E}' = \mathbf{J}\mathcal{E}.$

The electric field represented by this vector hits an antenna of our interferometer that has two feeds, each one sensitive to a specific polarization state of the vector in case of a perpendicularly incident electric field.
Let us assume that the $p$ and $q$ feeds are sensitive to the $x$ and $y$ polarization states of the signal respectively.
The feeds convert the respective electric fields into voltages and this conversion can be expressed as yet another Jones matrix yielding
\begin{equation}
 \mathcal{V} = \mathbf{J}'\mathcal{E}' \Rightarrow \begin{pmatrix}
v_p \\
v_q
\end{pmatrix} = \mathbf{J}'\begin{pmatrix}
e_x\\
e_y
\end{pmatrix}.\label{eq:v2e}
\end{equation}
Let us denote this antenna as $a$ and assume that there is another antenna in our interferometer denoted by $b$.
Voltages from each antenna are fed to a correlator that cross-correlates them to create 4 pairwise correlations that can be written as a $2\times 2$ matrix, known as the \textit{visibility matrix},
\begin{equation}
\mathbf{V}_{ab} = \langle \mathcal{V}_a\mathcal{V}_b^H \rangle = \begin{pmatrix}
\langle v_{ap}v_{bp}^* \rangle & \langle v_{ap}v_{bq}^* \rangle \\
\langle v_{aq}v_{bp}^* \rangle & \langle v_{aq}v_{bq}^* \rangle
\end{pmatrix} = \begin{pmatrix}
V_{pp} & V_{pq} \\
V_{qp} & V_{qq}
\end{pmatrix}
\label{eq:volt}
\end{equation}
which is related to the electric field correlations according to Eq. \ref{eq:v2e}, i.e.
\begin{align}
\mathbf{V}_{ab} = \mathbf{J}_a \begin{pmatrix}
\langle e_{x}e_{x}^* \rangle & \langle e_{x}e_{y}^* \rangle \\
\langle e_{y}e_{x}^* \rangle & \langle e_{y}e_{y}^* \rangle
\end{pmatrix} \mathbf{J}_b^H.
\label{eq:vmatrix}
\end{align}
Here $*$ denotes a complex conjugate, $H$ the conjugate transpose or Hermitian conjugate and $\langle \rangle$ the time averages.
Polarized waves are best described by Stokes parameters and their relation with the correlations of the electric field components, for a linear experiment, can be written as \citep{ha96}
\begin{equation}
 \begin{pmatrix}
  \langle e_{x}e_{x}^* \rangle & \langle e_{x}e_{y}^* \rangle \\
  \langle e_{y}e_{x}^* \rangle & \langle e_{y}e_{y}^* \rangle
 \end{pmatrix} = \begin{pmatrix}
  I+Q & U+iV \\
  U-iV & I-Q
 \end{pmatrix} \equiv \mathbf{B}
\end{equation}
where $\mathbf{B}$ is the \textit{brightness matrix}.
Therefore, Eq. \ref{eq:vmatrix} becomes
\begin{equation}
 \mathbf{V}_{ab}=\mathbf{J}_a \mathbf{B} \mathbf{J}_b^H
 \label{eq:rime1}
\end{equation}
which contains all effects along the signal path in the form of Jones matrices.
The effect fundamental to all interferometers is the phase difference between the measured voltages $\mathcal{V}_a$ and $\mathcal{V}_b$.
To account for the phase delays in Eq. \ref{eq:rime1}, consider the interferometer to be situated in a Cartesian coordinate system represented by $u,v,w$ and the antenna $a$ to be located at the coordinates $\mathcal{U}_a=(u_a,v_a,w_a)$.
The phase delay between the baselines $a$ and $b$ then becomes
\begin{equation}
 K_{ab} = e^{-2\pi i(u_{ab}l+v_{ab}m+w_{ab}(n-1))}
\label{eq:phase}
\end{equation}
where $\mathcal{U}_{ab} = \mathcal{U}_a-\mathcal{U}_b$; $l,m$ are the cosines of the right ascension and declination of the source respectively; and $n=\sqrt{1-l^2-m^2}$.
If we take out the phase delay scalar matrices ($K$-Jones) from $\mathbf{J}$ for both antennae and express them as a single scalar associated with the baseline, then Eq. \ref{eq:rime1} becomes
\begin{equation}
 \mathbf{V}_{ab} = \mathbf{J}_a \mathbf{B}K_{ab} \mathbf{J}_b^H = \mathbf{J}_a \mathbf{X}_{ab} \mathbf{J}_b^H
 \label{eq:rimesingle}
\end{equation}
where $\mathbf{X}_{ab}=\mathbf{B} K_{ab}$ is called the \textit{coherency matrix} as it represents the \textit{spatial coherence function} \citep{cl} of the electric field for this particular baseline.

If, instead of a single source, we have a continuum of sources, the visibility matrix has to be written as an integration over all directions within the field of view and the cumulative Jones matrix has to be separated into two different matrices, one representing the direction independent effects (DIE, $G$-Jones) and another the direction dependent effects (DDE, $E$-Jones),
\begin{align}
\mathbf{V}_{ab} &= \mathbf{G}_{a} \left[ \iint\limits_{l,m} \mathbf{E}_{a} \mathbf{B} K_{ab} \mathbf{E}_{b}^H \frac{dl dm}{n} \right] \mathbf{G}_{b}^H.
\label{eq:rime}
\end{align}
This is the standard equation to describe the mathematical model of a radio interferometer that, from now on, we will refer to as the \emph{measurement equation}.

\subsubsection{Mueller formalism} \label{s:mueller}
For understanding the effects of systematic errors on the images produced from the visibilities, it helps to write this equation in terms of baseline-based Mueller matrices ($\mathbf{M}$) instead of antenna-based Jones matrices ($\mathbf{J}$) remembering the relation between the two \citep{ha96},
\begin{equation}
\mathbf{M}_{ab} = \mathbf{S}^{-1} (\mathbf{J}_a \otimes \mathbf{J}_b^H) \mathbf{S}
\label{eq:m2j}
\end{equation}
where the coordinate transformation matrix,
\begin{equation}
\mathbf{S} = \frac{1}{2}\begin{pmatrix}
1 & 1 & 0 & 0 \\
0 & 0 & 1 & i \\
0 & 0 & 1 & -i \\
1 & -1 & 0 & 0 \\
\end{pmatrix}
\end{equation}
and $\otimes$ denotes the Kronecker product.
To do so, instead of taking the matrix product of $\mathbf{v}_a$ and $\mathbf{v}_b$ like in Eq. \ref{eq:volt}, we have to take their Kronecker product to get the voltage correlation vector $\mathcal{V}_{ab} = (V_{pp} \ V_{pq} \ V_{qp} \ V_{qq})^T$ where $T$ represents transpose.
Then Eq. \ref{eq:rime} becomes
\begin{equation}
\mathcal{V}_{ab} = \mathbf{G}_{ab} \iint\limits_{l,m} \mathbf{E}_{ab} \mathbf{S} \mathcal{I} K_{ab} \frac{dl dm}{n}
\label{eq:riMe}
\end{equation}
where $\mathbf{G}_{ab} = \mathbf{G}_{a} \otimes \mathbf{G}_{b}^H$, $\mathbf{E}_{ab} = \mathbf{E}_{a} \otimes \mathbf{E}_{b}^H$ and brightness vector $\mathcal{I} = (I \ Q \ U \ V)^T$.

\subsubsection{Stokes visibilities}
In order to describe the relation between Stokes parameters and voltage correlations in Fourier space, let us define
\begin{equation}
 V_Z^{(ab)} = \mathbf{J}_a Z K_{ab} \mathbf{J}_b^H
 \label{eq:stokesvis}
\end{equation}
where $V_Z^{(ab)}=V_I, V_Q, V_U, V_V$ is a \textit{Stokes visibility} and $Z=I,Q,U,V$ is a Stokes parameter.
Comparing equations \ref{eq:stokesvis}, \ref{eq:rimesingle} and \ref{eq:volt}, and remembering the definition of the coherency and brightness matrices, we can establish the relation between Stokes visibilities and the voltage correlations as \citep{sa,bu},
\begin{subequations} \label{eq:stokesV}
\begin{align}
& V_I = \frac{1}{2} (V_{pp}+V_{qq}) \\
& V_Q = \frac{1}{2} (V_{pp}-V_{qq}) \\
& V_U = \frac{1}{2} (V_{pq}+V_{qp}) \\
& V_V = \frac{1}{2i} (V_{pq}-V_{qp}).
\end{align}
\end{subequations}
\subsection{Systematic effects}\label{s:syseff}
\begin{figure}
\centering
\includegraphics[width=\linewidth]{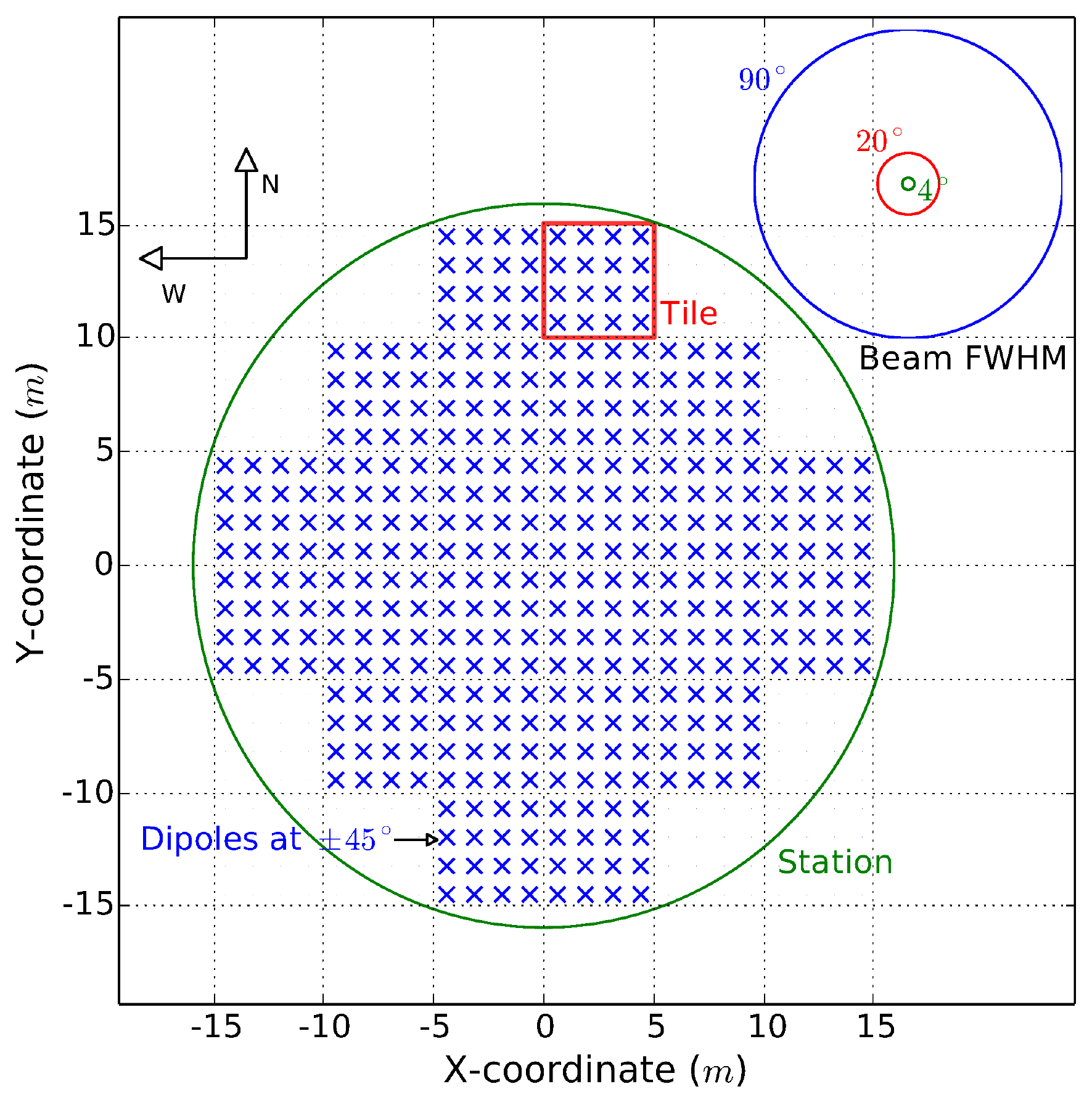}
\caption{Schematic diagram of a 24-tile LOFAR HBA station. A tile is made of 16 dual polarization dipoles. Dipoles see almost the whole sky (FWHM$\sim 90^\circ$), while the FWHM of a tile beam is $\sim 20^\circ$ and that of a station beam is only $\sim 4^\circ$.
There is a 15 cm gap between the tiles which is not shown here.}
\label{f:station}
\end{figure}

In this section, we will discuss the effects of the systematic errors ($G$ and $E$ Jones) on the Stokes visibilities and the Stokes parameters for the case of LOFAR, although the aforementioned formalism is universal.
LOFAR is a phased array covering the frequency range from 10--240 MHz.
LOFAR stations consist of two types of antennae--- LBA (low band antenna; 10--90 MHz) and HBA (high band antenna; 110--240 MHz).
We use the HBA stations in our simulations and a schematic diagram of a typical 24-tile LOFAR HBA core (situated within the central 3.5 km) station is shown in Fig. \ref{f:station}.
In this case, 16 dipoles are combined to create a tile and 24 tiles are combined to create a station \citep[for details see][]{vh}.

\subsubsection{Direction independent effects}\label{s:die}
To simplify calculations, while discussing DIEs, we will ignore the DDEs by assuming the $E$-Jones terms of Eq. \ref{eq:riMe} to be identity matrices.
Consequently, the Mueller-matrix form of the measurement equation (Eq. \ref{eq:riMe}) becomes,
\begin{equation}
\mathcal{V}_{ab} = \mathbf{G}_{ab} \iint\limits_{l,m} \mathbf{S} \mathcal{I} K_{ab} \frac{dl dm}{n} = \mathbf{G}_{ab} \iint\limits_{l,m} \mathbf{S} \widehat{\mathcal{V}}_Z \frac{dl dm}{n}
\label{eq:riMe1}
\end{equation}
where $\widehat{\mathcal{V}}_Z=\mathcal{I} K_{ab}$ represents the Stokes visibilities without any systematic errors.
The DIEs, denoted here by $\mathbf{G}_{ab}$, are caused by errors in the electronic gains of the antennae (gain errors) and non-orthogonal and/or rotated feeds (feed errors). Gain and feed errors, for antenna $a$, can be modelled by the Jones matrices,
\begin{equation}
\mathbf{G}_a^g = \begin{pmatrix}
g_{ap} & 0 \\
0 & g_{aq}
\end{pmatrix} \text{ \ and \ } 
\mathbf{G}_a^f = \begin{pmatrix}
1 & \epsilon_{ap} \\
-\epsilon_{aq} & 1
\end{pmatrix}
\end{equation}
where $g_{ap}$ is the gain error of the feed $p$ of the antenna $a$ and $\epsilon_{ap}$ is the spurious sensitivity of the $p$ feed to the $y$ polarization.
The Jones matrix for all DIEs, i.e. $G$-Jones of Eq. \ref{eq:rime}, then becomes $\mathbf{G}_{a}=\mathbf{G}_a^g \mathbf{G}_a^f$.
Gain and feed errors affect different Stokes visibilities (Eq. \ref{eq:stokesV}) in different ways which can be illustrated by taking into consideration how the Stokes visibilities observed by an instrument with DIEs differ from that of an error-free ideal instrument.
Let's assume that both $\mathbf{G}^g$ and $\mathbf{G}^f$ of the ideal instrument are identity matrices and for a realistic instrument gains and feeds are in error by,
\begin{equation}
\Delta \mathbf{G}_a^g = \begin{pmatrix}
\Delta g_{ap} & 0 \\
0 & \Delta g_{aq}
\end{pmatrix} \text{ \ and \ \ } \Delta \mathbf{G}_a^f = \begin{pmatrix}
0 & \epsilon_{ap} \\
-\epsilon_{aq} & 0
\end{pmatrix}.
\label{eq:delG}
\end{equation}
Then, seven error parameters (hereafter \textit{DI-error parameters}) can be defined following \citet[equations 36-42]{sa} as,
\begin{subequations}\label{eq:diep}
\begin{align}
&\delta_s = (\Delta g_{ap}+\Delta g_{aq}) + (\Delta g_{bp}^*+\Delta g_{bq}^*) &\\
&\delta_{I,Q} = (\Delta g_{ap}-\Delta g_{aq}) + (\Delta g_{bp}^*-\Delta g_{bq}^*) &\\
&\delta_{U,V} = (\Delta g_{ap}-\Delta g_{aq}) - (\Delta g_{bp}^*-\Delta g_{bq}^*) &\\
&\delta_{Q,U} = (\epsilon_{ap}+\epsilon_{aq}) + (\epsilon_{bp}^*+\epsilon_{bq}^*) &\\
&\delta_{I,U} = (\epsilon_{ap}-\epsilon_{aq}) + (\epsilon_{bp}^*-\epsilon_{bq}^*) &\\
&\delta_{I,V} = (\epsilon_{ap}+\epsilon_{aq}) - (\epsilon_{bp}^*+\epsilon_{bq}^*) &\\
&\delta_{Q,V} = (\epsilon_{ap}-\epsilon_{aq}) - (\epsilon_{bp}^*-\epsilon_{bq}^*) &
\end{align}
\end{subequations}
where the subscript $I,Q$ stands for mixing between Stokes $I$ and $Q$.
Now, if the difference between the ideal Stokes visibilities and the Stokes visibilities affected by these errors is $\Delta \mathcal{V}= \mathcal{V}_{ab}^{\rm ideal} - \mathcal{V}_{ab}$, then by assuming errors to be very small it can be shown that \citep[see][appendix B]{sa},
\begin{equation} \label{eq:dieM}
\Delta \mathcal{V} = -\frac{1}{2}\begin{pmatrix}
\delta_{s} & \delta_{I,Q} & \delta_{I,U} & -i\delta_{I,V} \\
\delta_{I,Q} & \delta_{s} & \delta_{Q,U} & -i\delta_{Q,V} \\
\delta_{I,U} & -\delta_{Q,U} & \delta_{s} & i\delta_{U,V} \\
-i\delta_{I,V} & \delta_{Q,V} & -i\delta_{U,V} & \delta_{s} \\
\end{pmatrix} \mathcal{V}_{ab}^{\rm ideal}.
\end{equation}
Here, the $4\times 4$ matrix is the instrumental Mueller matrix for the DIEs (hereafter \textit{DI-Mueller}) and it determines the full Stokes response of an instrument without any direction dependent errors.
It can be seen from the equation that a completely unpolarized source ($Q,U,V=0$) will appear to have non-zero Stokes $Q$, $U$ and $V$ in an interferometric observation because of the DIEs $\delta_{I,Q}$, $\delta_{I,U}$ and $\delta_{I,V}$ respectively and these same errors will cause leakage into Stokes $I$ from Stokes $Q$, $U$ and $V$ respectively.
The DIE-parameters can be used to determine calibration errors if, instead of comparing the ideal and the actual gains, we compare the input and the solved gains \citep{sa}.

\subsubsection{Direction dependent effects}\label{s:dde}
\begin{figure*}
\begin{minipage}[b]{0.62\linewidth}
\centering
\includegraphics[width=\textwidth]{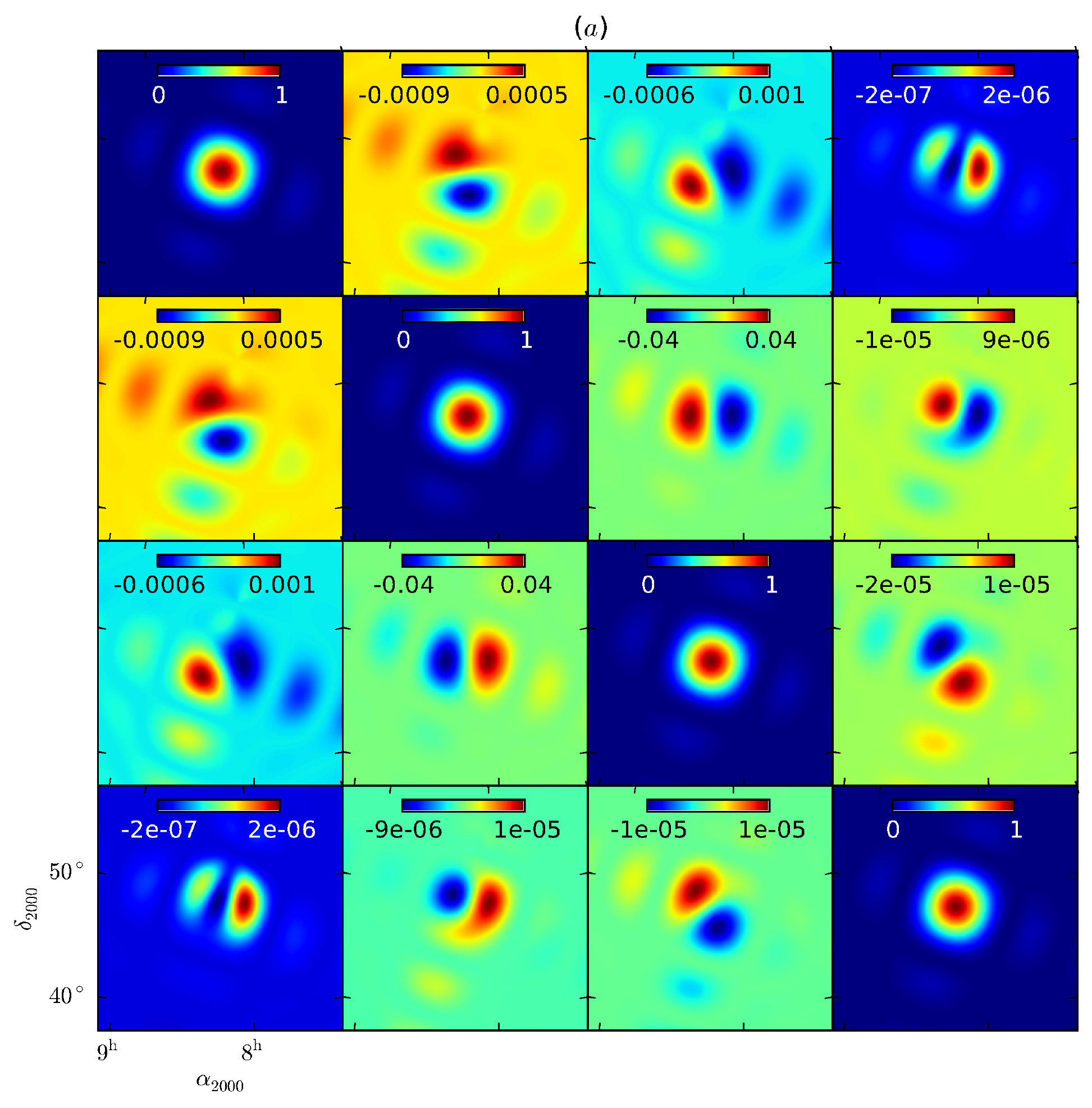}
\end{minipage}
\begin{minipage}[b]{0.3\linewidth}
\centering
\includegraphics[width=\textwidth]{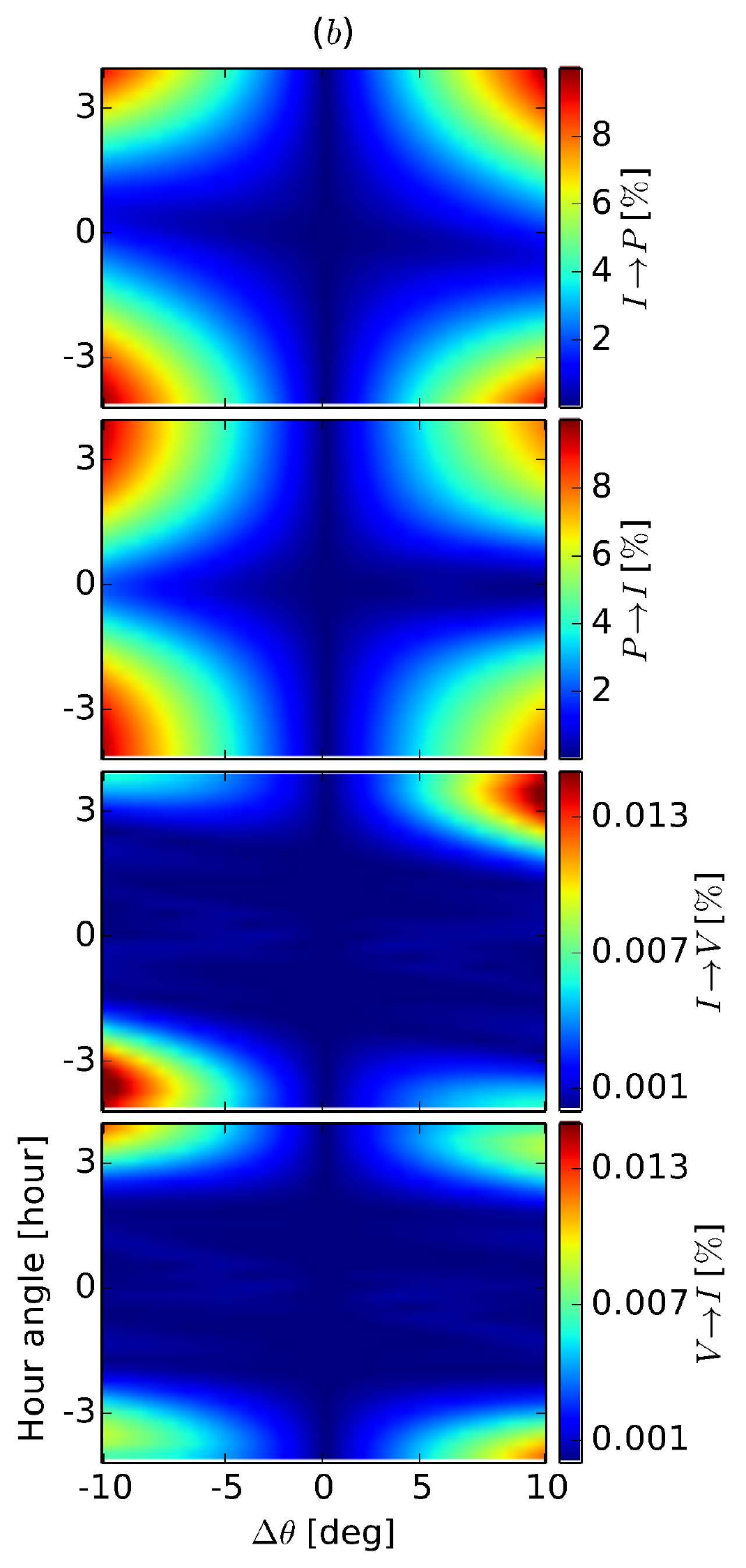}
\end{minipage}
\caption{(\textit{a}) Direction dependent Mueller matrix representing the polarization response of the baseline 0-1 (127 $m$) of LOFAR at 150 MHz over the 3C196 field ($20^\circ \times 20^\circ$) at the time when the centre of the field culminates.
(\textit{b}) Spatio-temporal profiles as a percentage of total intensity---i.e. first row, first column ($M_{11}$) of the matrix representing Stokes $I$---for leakages from (1) $I$ to linear polarization ($P$), i.e. $\sqrt{M_{12}^2+M_{13}^2}$; (2) linear to $I$, i.e. $\sqrt{M_{21}^2+M_{31}^2}$; (3) $I$ to circular, $M_{14}$ and (4) circular to $I$, $M_{41}$.
Here, $\Delta\theta$ represents distance from the phase centre.
See section \ref{s:dde} for details.}
\label{f:mueller}
\end{figure*}
Direction dependent errors in a radio interferometer are caused mainly by the Earth's ionosphere and the primary beams---i.e. the radiation patterns---of the antennae.
Here, we restrict ourself only to the LOFAR beam errors.
The beam we use for the bowtie dipoles has been modelled by an analytic expression whose coefficients are determined by fitting to a numerically simulated beam raster generated by the ASTRON Antenna Group (\citealp{ha11}; hereafter H11).
Here, we will give a brief overview of this model; for further details we refer the readers to H11.

From basic symmetry considerations a generic expression for a dual dipole antenna $E$-Jones matrix has been derived by H11 which, for azimuth $\phi$ and zenith angle $\theta\equiv(\pi/2-$ elevation) can be written as,
\begin{equation}
\mathbf{E}_e(\theta,\phi) = \sum\limits_{k'=0}^N \mathbf{R}(k',\phi) \mathbf{P}_k(\theta)
\label{eq:dbeam}
\end{equation}
where the azimuth dependent \emph{rotation matrix}
\begin{align}
\mathbf{R}(k',\phi) = \begin{pmatrix}
\cos [(-1)^{k'} (2k'+1) \phi] & -\sin [(-1)^{k'} (2k'+1) \phi] \\
\sin [(-1)^{k'} (2k'+1) \phi] & \cos [(-1)^{k'} (2k'+1) \phi]
\end{pmatrix}
\end{align}
and the zenith angle and frequency ($\nu$) dependent \emph{projection matrix} that contains the detailed geometry of the dipoles and the ground plane is
\begin{align}
\mathbf{P}_{k'}(\theta,\nu) = \begin{pmatrix}
p_{\theta,k'}(\theta,\nu) & 0 \\
0 & -p_{\phi,k'}(\theta,\nu)
\end{pmatrix},
\end{align}
and $k'=0$ gives the `ideal' beam, whereas the higher order terms represent the differences between the ideal and the more realistic beams.
Each element of the projection matrix $p(\theta,\nu)$, for each harmonic $k'$, is calculated as $\bar{\theta} [\mathbf{C}] \bar{\nu}$ where $\bar{\theta}$ is a row vector $(\theta^0 \ \theta^1 \ ... \ \theta^{N_\theta})$, $\bar{\nu}$ is a column vector $(\nu^0 \ \nu^1 \ ... \ \nu^{N_\nu})^T$, $[\mathbf{C}]$ is a 2D matrix of dimensions $(N_\theta+1)\times (N_\nu+1)$ that contains the complex coefficients determined by fitting to an electromagnetic simulation, and $N_\theta=N_\nu=4$.

In Eq. \ref{eq:dbeam}, $\mathbf{E}_e$ has been expressed in a topocentric (azimuth-zenith angle) coordinate, but in reality the source is carried around through the beam by the apparent rotation of the sky during an observation.
To account for this effect, the position of the source is transformed from equatorial celestial coordinate system to the topocentric system.
For polarized sources, there is an additional factor--- the relative rotation between the equatorial and the topocentric grids at the position of the source that causes the beam to rotate with the parallactic angle, known as the \textit{parallactic rotation} which has been incorporated in the dipole beam model as a separate Jones matrix.
Hereafter, by $\mathbf{E}_e$ we will refer to an element beam where all these effects have been taken into account.

\begin{figure*}
\begin{minipage}[b]{0.37\linewidth}
\includegraphics[width=\textwidth]{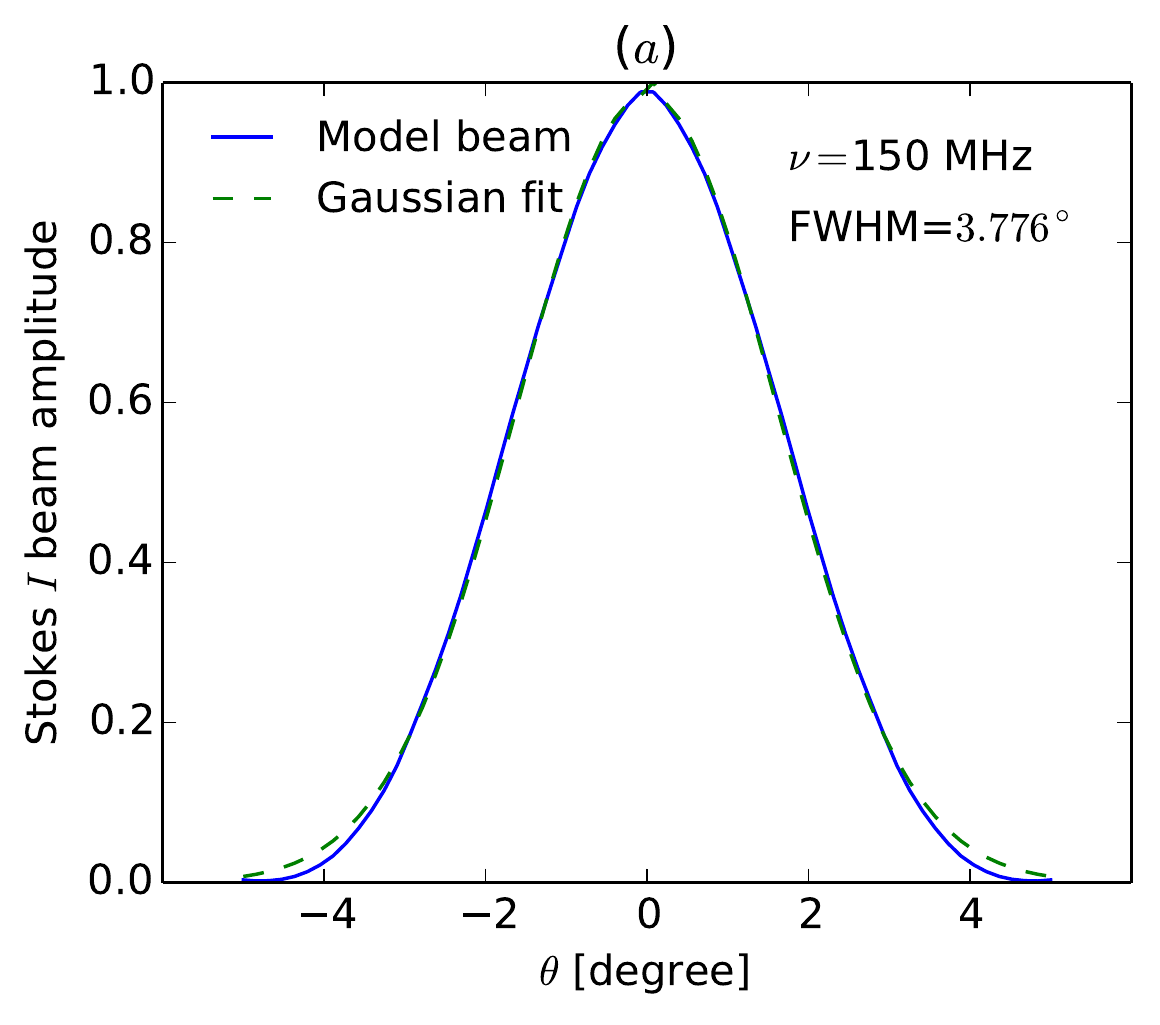}
\end{minipage}
\begin{minipage}[b]{0.38\linewidth}
\includegraphics[width=\textwidth]{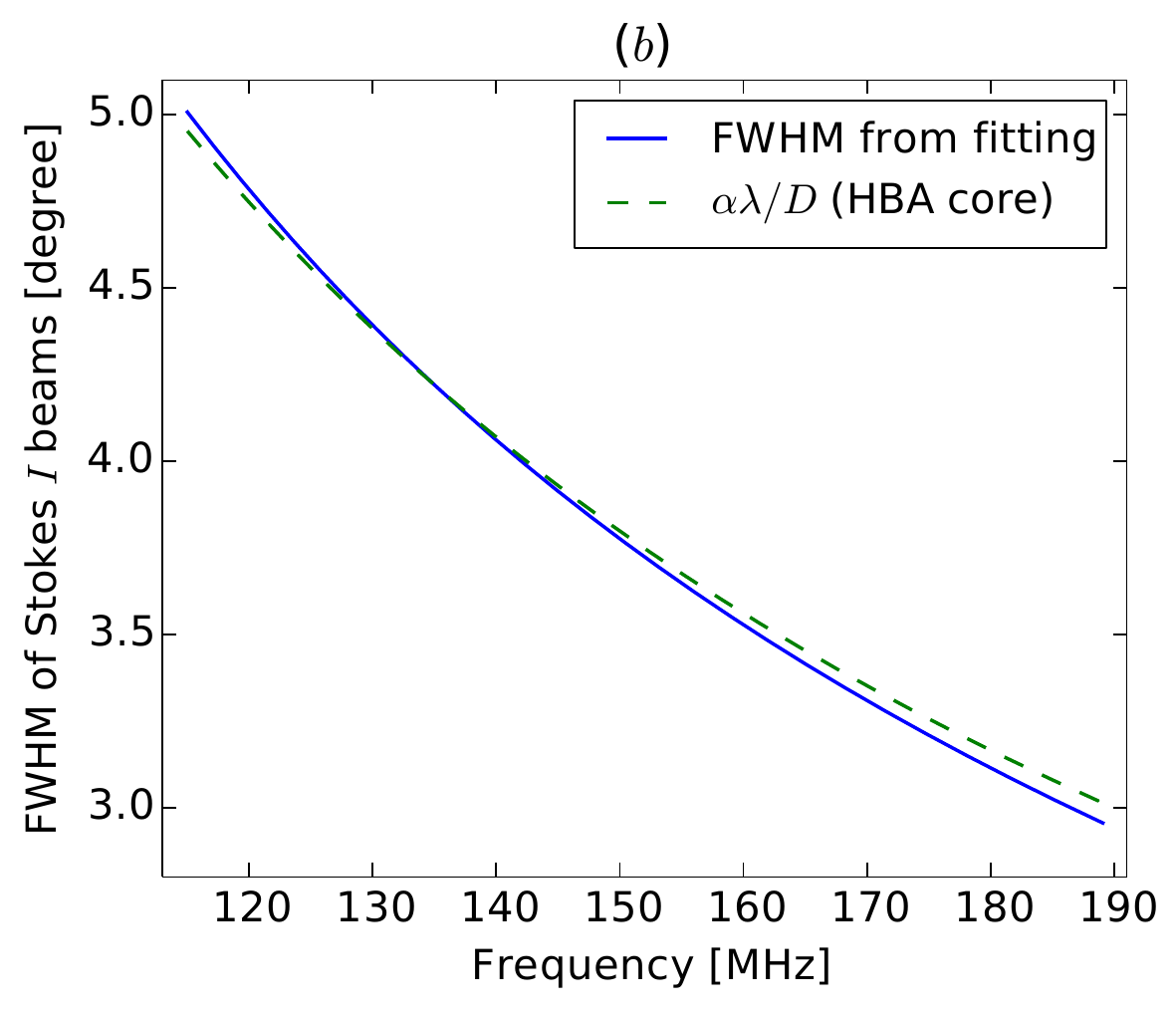}
\end{minipage}
\begin{minipage}[b]{0.22\linewidth}
\includegraphics[width=\textwidth]{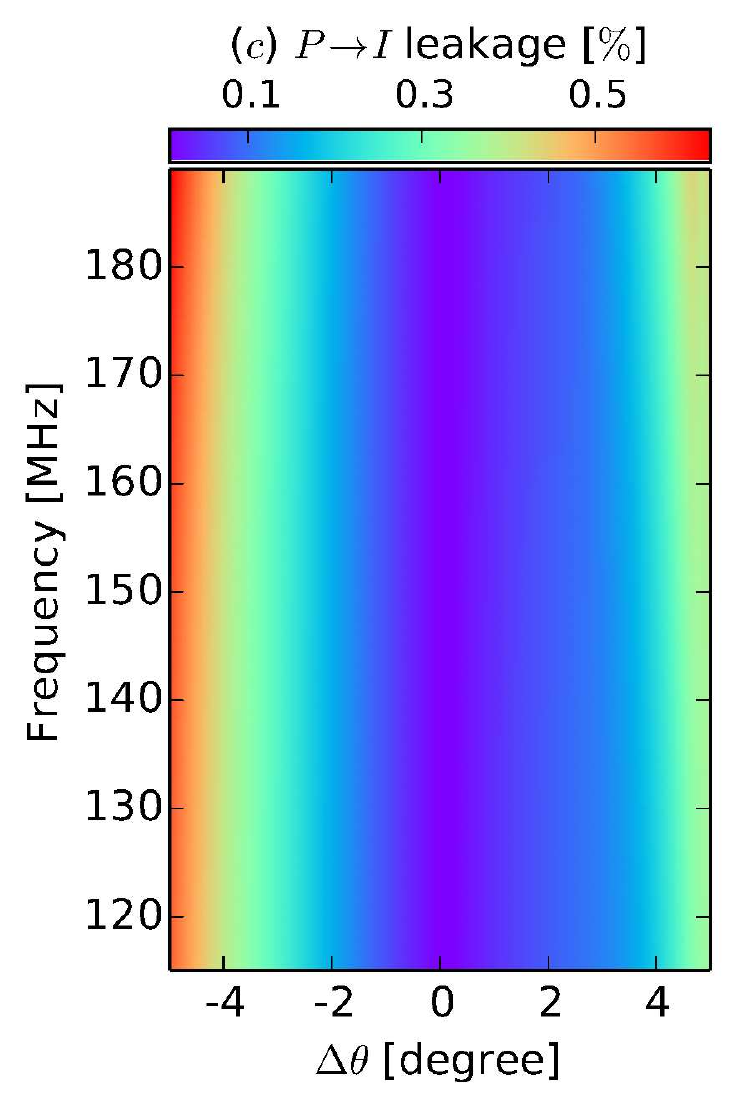}
\end{minipage}
\caption{(\textit{a}) Gaussian fit to the azimuthally averaged Stokes $I$ response of the 0-1 baseline of LOFAR at 150 MHz over the 3C196 field when the field culminates ($M_{11}$ component of Fig. \ref{f:mueller}a). (\textit{b}) FWHM of the Stokes $I$ beam at different frequencies (solid); the $\alpha\lambda/D$ curve (dashed) is overplotted. (\textit{c}) A single line through the centre of the Mueller term responsible for linear polarization leakage (see caption of Fig. \ref{f:mueller}) at different frequencies. The leakage is shown as a percentage of Stokes $I$ flux density.}
\label{f:freq}
\end{figure*}

In an element beam Jones matrix the diagonal terms determine the primary beam of the element and the off-diagonal terms the level of cross-polarization.
Errors related to antenna pointing, beamwidth and beam ellipticity are all included in the diagonal terms.
For a dipole of size $D\sim 1.25$ m the FWHM at 150 MHz becomes $\lambda/D\sim 90^\circ$ and the shape of the diagonal terms of the matrix is similar to an Airy pattern.
The polarization response of a LOFAR station is completely determined by $\mathbf{E}_e$.
Therefore, it would be interesting to analyse the beam Mueller matrix corresponding to an interferometer constructed by two such elements before entering into the discussion of the tile and the station beams.

In a two-element interferometer, the component at the first row and first column of the Mueller matrix (hereafter $M_{11}$) represents the Stokes $I$ response of the interferometer to a completely unpolarized point source of unity flux and $M_{12}$ gives the corresponding Stokes $Q$ response.
Examples of Stokes $I$ and $Q$ responses of a LOFAR LBA dipole can be seen in Fig. 3.8 and 3.9 of \citet[][hereafter B12]{br12} respectively.
From the figures we see that Stokes $I$ response is almost circular with amplitudes decreasing from the centre toward the edges until the first null.
Stokes $Q$ response, on the other hand, has a cloverleaf pattern with $2$-fold symmetry corresponding to the physical structure of the dual dipole.
The cross-polarization over a beam is conventionally measured by the ratios $Q(\theta)/I(\theta)$, $U(\theta)/I(\theta)$ and $V(\theta)/I(\theta)$.
Comparing Fig. 3.9 and 3.8, B12 finds that $Q/I$ is lowest at the centre and increases quadratically with $\theta$ and reaches a value of $0.5$ at the FWHM.
It implies that an unpolarized source situated at FWHM of a dipole beam will become 50\% polarized in the observed data due to instrumental polarization.

The beams of the 16 dipoles ($\mathbf{E}_e$) in a tile are combined in an analogue way to form the tile beam which is narrower ($\sim 20^\circ$, Fig. \ref{f:station}) and the beams of all the tiles in a station are digitally combined to form the station beam which has the smallest width ($\sim 4^\circ$).
Assuming the tile beams ($\mathbf{E}_t$) have been created by phasing the constituting dipole beams, the beam of the station $a$ can be written as \citep{ya9}
\begin{equation}
\mathbf{E}_a(\theta,\phi) = \mathbf{w}^H \mathbf{v}(\mathbf{k}) \odot \mathbf{E}_t(\theta,\phi)
\label{eq:beam}
\end{equation}
where $\odot$ denotes the Hadamard product, $\mathbf{k}$ is the wave vector, $\mathbf{v}(\mathbf{k})$ is the steering vector, i.e. the delay an incoming wavefront experiences depending on the position ($\mathbf{r}_i$) of the observing tile in a station that can be expressed as
\begin{equation}
\mathbf{v}(\mathbf{k}) = \begin{pmatrix}
e^{-j\mathbf{k}.\mathbf{r}_0} \\
e^{-j\mathbf{k}.\mathbf{r}_1} \\
\vdots \\
e^{-j\mathbf{k}.\mathbf{r}_{N-1}}
\end{pmatrix}
\end{equation}
for $N$ number of tiles and $\mathbf{w}$ is the weight vector that contains the complex weights associated with each tile.
Station beams cut only a small portion of the element beam and get a polarization response depending on which part of the element beam it is tracing.
The sidelobes of the station beam cut yet another part of the element beam and accordingly acquire a different polarization response.
Station beams that are formed to track a source in the sky follow a trace in azimuth and elevation over the polarized element beam.
Hereafter, by \emph{beam} we will refer to the beam of a single station, $\mathbf{E}_a$.

We could, in principle, derive a direction dependent equivalent of Eq. \ref{eq:dieM} using $\mathbf{E}_a$ as the only systematic error and ignoring the DIEs, but it will be much more complicated in this case.
So, instead, we numerically calculate the baseline-dependent Mueller matrices (e.g. $\mathbf{E}_{ab}$) from the constituent station beams ($\mathbf{E}_a$ and $\mathbf{E}_b$) following the formalism of section \ref{s:mueller}.
Such a Mueller matrix for baseline 0-1 (a 127 $m$ baseline formed by the two sub-stations of the central core stations, CS001HBA0 and CS001HBA1) at 150 MHz, at the time when the centre of the target field ($20^\circ \times 20^\circ$) culminates has been shown in Fig. \ref{f:mueller}a.
The components of the matrix have been normalized with respect to the Mueller matrix at the phase centre resulting in a differential Mueller matrix; hereafter, by differential beam or nominal beam we will refer to this form of the Mueller matrix.
Let's denote this matrix by $\mathbf{M}^{01}$, where the superscript represents the station numbers.

$\mathbf{M}^{01}$ can be thought of as a direction dependent equivalent of the DI-Mueller (Eq. \ref{eq:dieM}), hence we can call it the \textit{DD-Mueller}.
By comparing these two matrices, we can see that $M_{21}$ component of the DD-Mueller will cause Stokes $I$ to leak into Stokes $Q$.
The off-diagonal terms of $\mathbf{M}^{01}$ show the spatial variation of the instrumental polarization--- it is lowest at the phase centre and increases toward the edges until the first null and then, after a gap, we get further polarization at the location of the first sidelobe.
In addition to the spatial variation, all components of the instrumental Mueller matrix also vary with zenith angle, or equivalently with hour angle, of the source during an observation.
To show the dependence on the directions and sidereal time simultaneously, i.e. spatio-temporal dependence, we calculated $\mathbf{M}^{01}$ for all hour angles.
In Fig. \ref{f:mueller}b, we show spatio-temporal profiles of various leakages as a percentage of total intensity.
Leakage from linear polarization to total intensity, i.e. $\sqrt{M_{12}^2+M_{13}^2}/M_{11}\times 100$, at different distances from the phase centre ($x$ axis) and at different hour angles ($y$ axis) during an eight-hour observation is shown in the top panel.
The second panel shows fractional leakage from Stokes $I$ to linear polarization and the third and fourth panels show fractional $I\rightarrow V$ and $V\rightarrow I$ leakages respectively.
These figures show the variation of the leakages along a single line through the centre of the field at every hour angle during a night-long observation.

From the spatio-temporal profiles, we see that leakage increases with both distance from the phase centre and zenith angle.
During the beginning and the end of the observation zenith angle is very high and the beam is extremely attenuated which results in a very high percentage of leakage.
Leakages vary across the FoV mainly due to polarization aberrations caused by geometric projection of the antenna on the plane perpendicular to the line of sight (see section 5.3 and Fig. 2 of \citealt{ca}).
The projection changes as a function of direction and zenith angle because of both the coordinate rotation and parallactic rotation that were introduced in the beam model (as discussed before).
We see that at high zenith angle the leakages change more rapidly, but these effects can be considered constant within ten minutes (B12) which is an useful assumption for primary beam correction.

Besides direction and elevation, the width and shape of the beam also vary with frequency.
Fig. \ref{f:freq}a shows a Gaussian fit to the azimuthally averaged station beam ($M_{11}$) that gives us an FWHM of $3.8^\circ$ at 150 MHz.
Fig. \ref{f:freq}b shows the beamwidths obtained by Gaussian fitting as a function of frequency and we can see that the curve closely follows the $\alpha\lambda/D$ relation where $\lambda$ and $D$ denote wavelength and station size respectively (for an analogous fitting, see Fig. 21 of \citealt{vh}).
Leakages also vary with frequency, albeit not in a very prominent way; as evident from Fig. \ref{f:freq}c, within approximately ten degrees leakage changes very slowly with frequency.
Therefore, if we have multi-frequency data, the leakages can be removed by utilizing their spectral smoothness.

Ideally, the beam should be exactly same for all elements and, consequently, for all baselines, for traditional calibration to work efficiently, but making them slightly different in configuration could be advantageous in another way.
In case of LOFAR, although all dipoles are rotated into the same position, station configurations are rotated with respect to one another to minimize blind angle effects and to average out the effect of grating lobes (B12).

\subsection{Calibration and imaging}
In DI-calibration, it is assumed that all baselines of an array observe the Fourier transform of a common sky which is only true if DDEs are taken to be identical across all antennae.
Consequently, $\mathbf{E}_a$ of Eq. \ref{eq:rime} becomes a function of just $l,m$ and the common sky observed by all baselines becomes $\mathbf{B}_c=\mathbf{E}\mathbf{B}\mathbf{E}^H$, i.e. the true sky attenuated by the beam.
Then, Eq. \ref{eq:rime} can be written as
\begin{align}
\mathbf{V}_{ab} &= \mathbf{G}_{a} \mathbf{X}^c_{ab} \mathbf{G}_{b}^H
\end{align}
where $\mathbf{X}^c_{ab}$ is the element by element 2D Fourier transform of $\mathbf{B}_c$.
The most widely used DI-calibration method, \textit{self-calibration} or selfcal works with this form of the measurement equation.
The first step of selfcal is to create a model of the observed sky and to `predict' the corresponding visibilities, $\mathbf{V}_{ab}^{\rm mod}$ that an interferometer would produce.
Then, the values of $\mathbf{G}$ terms that minimize $\mathbf{V}_{ab}^{\rm mod}-\mathbf{V}_{ab}$ are determined.
$\mathbf{G}$ terms can be calculated to a very high accuracy, because an array provides over-determined information as $N(N-1)$ complex visibilities are available for computing only $2N-2$ error parameters, $N$ being the number of antennae.

The inferred values ($\tilde{\mathbf{G}}$) are applied to the observed visibilities to yield the corrected visibilities as
\begin{align}
\mathbf{V}_{ab}^{\rm corr} &= \tilde{\mathbf{G}}_{a}^{-1} \mathbf{V}_{ab} \tilde{\mathbf{G}}_{b}^{-H}.
\label{eq:vcorr}
\end{align}
Inverse Fourier transform of the weighted and gridded visibilities produce a `dirty' image, which is the true sky convolved with the PSF.
To recover the true sky as sampled by the visibilities as closely as possible, the PSF is deconvolved from the dirty image iteratively producing a `clean' image. As the primary beam has not been corrected for, this clean image is actually the true sky attenuated by the primary beam ($\mathbf{B}_c$).
If the primary beam is assumed to be same for all antennae and at all times, the true brightness distribution $\mathbf{B}$ can be extracted from $\mathbf{B}_c$ by just multiplying it with the inverse of $\mathbf{E}$.
Traditionally, this is what has been done for dish instruments with small FoV.
But in case of wide FoV instruments, e.g. LOFAR, time-frequency-baseline variations of the instrumental Mueller matrices ($\mathbf{M}$, Fig. \ref{f:mueller}) cannot be ignored and one way of dealing with this is AW-projection \citep{ta}.

\subsubsection{AW-projection} \label{s:aw}
The problem of imaging can be expressed in Mueller formalism as $\mathbf{V} = \mathcal{A}\mathbf{I} + \mathbf{\epsilon}$ where $\mathbf{V}$ is the total set of visibilities, $\mathbf{I}$ is the set of Stokes images to be estimated, $\mathbf{\epsilon}$ is the noise, $\mathcal{A} = \mathcal{W} \mathcal{S} \mathcal{F} \mathcal{M}$ ignoring the ionospheric effects, $\mathcal{W}$ is the set of visibility weights, $\mathcal{S}$ is the sampling function, $\mathcal{F}$ is the Fourier transform kernel, and $\mathcal{M}$ is the Mueller matrix corresponding to the primary beam.
Each of these parameters is a multi-dimensional matrix \citep[for explanation see][]{ta}.
AW-projection, as implemented in {\tt AWImager}, calculates $\hat{\mathbf{I}}$, an estimate of $\mathbf{I}$, iteratively as,
\begin{equation}
\hat{\mathbf{I}}^{n+1} = \hat{\mathbf{I}}^n + \mathbf{\Phi}
\mathcal{A}^H(\mathbf{V}-\mathcal{A}\hat{\mathbf{I}}^n)
\end{equation}
where $\mathbf{\Phi}$ is a non-linear operator that estimates the deconvolved sky from the residual dirty image $\mathcal{A}^H(\mathbf{V} - \mathcal{A}\hat{\mathbf{I}}^n)$.
Here the construction of the residual dirty image constitutes the major cycle and the deconvolution the minor cycle.
Note that $\mathcal{A}\hat{\mathbf{I}}^n$ is the forward Fourier transform taking into account all instrumental effects and this has to be done accurately for the solutions to converge; during prediction of visibilities using {\tt AWImager}, only this step is performed.
On the other hand, during minor cycle only an approximation of $(\mathcal{A}^H\mathcal{A})^{-1}$ is calculated and applied on the residual.
A-projection, as described in \citet{bh08}, is a fast way for applying $\mathcal{A}$ or $\mathcal{A}^H$.
In {\tt AWImager}, the element beam ($\mathbf{E}_e$) and the array factor ($\mathbf{w}^H\mathbf{v}(k)$ of Eq. \ref{eq:beam}) of LOFAR have been taken out of the $\mathcal{M}$ matrix of the A-term and they are applied separately.

\subsection{Flux conversion} \label{s:conv}
For easier comparison with the predicted level of the EoR signal we convert fluxes to intensities and express them as temperature.
If $F_{Jy}$ is the flux of a radio source in Jy, then the corresponding intensity in K units can be written as,
\begin{equation}
T_K = \frac{\lambda^2 F_{Jy}}{2k_B\Omega_E} 10^{-26}
\label{eq:conv}
\end{equation}
where $k_B$ is the Boltzmann constant and $\Omega_E=\pi\theta^2/(4\ln2)$ is the beam solid angle, $\theta$ being the FWHM of the Gaussian restored PSF calculated during imaging.

\subsection{Rotation measure synthesis} \label{s:rm}
The rotation of the plane of polarization ($\chi$) of a linearly polarized signal while propagating through a magnetized plasma is called Faraday rotation which, for a single Faraday screen along the LOS, can be written mathematically as $\chi=\chi_0+\Phi\lambda^2$ where $\chi_0$ is the intrinsic polarization angle and Faraday depth,
\begin{equation}
\Phi = 0.81 \int_{\rm source}^{\rm observer} n_e B_\parallel dl
\end{equation}
where $n_e$ is the density of electrons and $B_\parallel$ is the magnetic field component along the LOS.
Note that rotation measure (RM) is defined as $d\chi/d\lambda^2$ and hence for a single phase screen along the LOS it is equivalent to Faraday depth.
Polarized surface brightness per unit Faraday depth, $F(\Phi)$ can be obtained from the polarized surface brightness per unit squared-wavelength, $P(\lambda^2)$ using the technique of RM-synthesis \citep{br}; mathematically,
\begin{equation}
F(\Phi) = R(\Phi) \star \int_{-\infty}^\infty P(\lambda^2) e^{-2i\Phi\lambda^2} d\lambda^2
\end{equation}
where $R(\Phi)$ is the Fourier transform of the wavelength sampling function, known as `rotation measure spread function' (RMSF) and $\star$ denotes convolution.

The polarized brightness, $\mathcal{P}=Q+iU$\footnote{In this paper $\mathcal{P}$ always refers to $Q+iU$, while $P$ is always $|Q+iU|$, and note that the 2D and 3D power spectra, denoted by $P_{2D}$ and $P_{3D}$ respectively, are not related to $\mathcal{P}$ or $P$.} is a complex valued function and, hence, $F(\Phi)$ is also complex.
However, a Faraday dispersion function for real valued Stokes $I$, $F_I(\Phi)$ can also be calculated assuming its imaginary parts to be zero in all spectral bands \citep[e.g.][]{ge}.
As the Fourier transform of a real function is always Hermitian, $F_I^*(\Phi)=F_I(-\Phi)$.
The same can be done for Stokes $V$.
In section \ref{s:gal}, we will present some of our results in terms of $F(\Phi)$, $F_I(\Phi)$ and $F_V(\Phi)$.

\subsection{Power spectrum analysis}
The power spectrum (hereafter PS) of an image is the measure of the variance per unit angular wavenumber ($k=2\pi/\theta$).
As the first detections of the EoR signal will be statistical, and its PS is the most widely used statistic \citep[e.g.][]{bow,ha10,mo,pa14,ch14}, most of our analysis will be done through PS.
We present three types of PS: 2D, 3D cylindrical and 3D spherical, and in all of them the wavenumbers are converted to the unit of comoving Mpc$^{-1}$ at the redshift corresponding to the observing frequency.
PS can be calculated from the weighted visibilities directly.
As the imaging process puts weights on the visibilities and calculates the resulting PSF, we have measured the PS from the Fourier transform (hereafter FT) of the images remembering that the squared complex modulus of a FT yields the PS of a signal.

\subsubsection{2D power spectrum} \label{s:ps2d}
Assume that $\breve{\mathbf{I}}_{uv}$ is the 2D FT of the image $\mathbf{I}_{lm}$ where $u,v$ represent the spatial frequencies corresponding to the angular scales $l,m$.
The minimum and maximum spatial frequencies of $\breve{\mathbf{I}}_{uv}$ are determined by $1/(N_x\theta_{\rm pix})$ and $1/(2\theta_{\rm pix})$ respectively where $\theta_{\rm pix}$ is the angular size of the pixels in $\mathbf{I}_{lm}$ and $N_x=\sqrt{N_l^2+N_m^2}$ where $N_l$ and $N_m$ are the total number of pixels in $l$ and $m$ directions respectively.
We cut the portion of $\breve{\mathbf{I}}_{uv}$ delimited by the minimum and maximum physical baselines and calculate the 2D PS as $P_{2D}(u,v)=|\breve{\mathbf{I}}_{uv}^{\rm cut}|^2$.

To produce 1D angular PS, we divide $P_{2D}$ in several concentric circular bins and calculate the average power at every bin.
Finally, we plot the average power in the bins as a function of comoving transverse wavenumbers corresponding to the bins defined as \citep[][equations 2-3]{mh},
\begin{equation} \label{eq:k_perp}
k_\perp=\frac{2\pi U_\lambda}{D_c(z)}
\end{equation}
where $U_\lambda=\sqrt{u^2+v^2}$ in units of wavelengths, transverse comoving distance at redshift $z$, $D_c(z)=\int_0^z dz'/E(z')$, and dimensionless Hubble parameter, $E(z)=[\Omega_m(1+z)^3+\Omega_k(1+z)^2+\Omega_\Lambda]^{1/2}$, $\Omega_m$, $\Omega_k$ and $\Omega_\Lambda$ being the matter density, curvature and cosmological constant parameters respectively.
Thus, we obtain $k_\perp$ in units of Mpc$^{-1}$ and $P_{2D}(k_\perp)$ in units of K$^2$ Mpc$^2$. Note that the minimum and maximum values of $k_\perp$ are determined by $U_\lambda^{\rm min}$ and $U_\lambda^{\rm max}$ respectively, as shown in \citet[][equations 13-14]{ve}.

\subsubsection{3D power spectrum} \label{s:ps3d}
Assume that $\breve{\mathbf{I}}_{uv\eta}$ is the 3D FT of the image $\mathbf{I}_{lm\nu}$ where $\eta$ represents the LOS spatial frequency corresponding to the LOS distance signified by the frequency $\nu$ \citep[see][Fig. 2]{mh}.
After taking only the portion of the cube that represents real baseline distribution as before, the 3D PS can be calculated as $P_{3D}(u,v,\eta)=|\breve{\mathbf{I}}_{uv\eta}^{\rm cut}|^2$.
Two types of binned PS can be calculated from this PS-cube: cylindrical, $P_{3D}(k_\perp,k_\parallel)$, and spherical, $P_{3D}(k)$.

In the cylindrical case, averaging is done in concentric cylindrical bins centred on the centre of the cube.
Hence, $P_{3D}(k_\perp,k_\parallel)$ is the average power of all $uv$ cells within a logarithmic cylindrical bin around $k_\perp,k_\parallel$ where the comoving LOS wavenumber,
\begin{equation} \label{eq:k_para}
k_\parallel = \eta \frac{2\pi H_0 E(z) \nu_{21}}{c(1+z)^2},
\end{equation}
$\nu_{21}$ being the rest frequency of 21-cm radiation emitted by HI, and $k_\perp$ is the same as defined by Eq. \ref{eq:k_perp}.
The minimum and maximum values of $k_\parallel$ are given by $\eta_{\rm min}=1/B$ and $\eta_{\rm max}=1/\Delta \nu$ respectively where $B$ is the bandwidth and $\Delta \nu$ is the frequency resolution provided by the instrument.
From the minimum and maximum values of $k_\perp$ and $k_\parallel$, it is evident that the boundaries of the $k$-space are defined by the instrumental parameters \citep[see e.g.][Fig. 4]{ve}.
Instead of showing the raw power we plot the quantity $\Delta^2(k_\perp,k_\parallel) = k_\perp^2 k_\parallel P_{3D}(k_\perp,k_\parallel)/ (2\pi)^2$ in our 2D figures which has the dimensions of temperature squared.

For constructing the spherical 3D PS, we divide the PS-cube in concentric spherical annuli around the centre of the cube and average the power in every annulus. Consequently, we get a 1D PS as a function of $k=\sqrt{k_\perp^2+k_\parallel^2}$. Here, we plot the quantity $\Delta^2(k)=k^3 P_{3D}(k)/(2\pi^2)$ that has the same dimensions as $\Delta^2(k_\perp,k_\parallel)$.

\section{Simulations of extragalactic foreground} \label{s:egal}
To show the effects of direction independent errors on calibration, we simulate the observations of a mock sky with point sources.
In case of the direction dependent errors, we first simulate a mock sky to show the trend of the effects, and then proceed to simulate the realistic sky to quantify the effects expected in the LOFAR-EoR observations.
We did not include any additive noise in the simulations described in this section.
Below we describe the general pipeline of the simulations followed by the set-ups and results of the specific simulations.

\subsection{Pipeline} \label{s:ppipe}
\begin{figure}
\includegraphics[width=\linewidth]{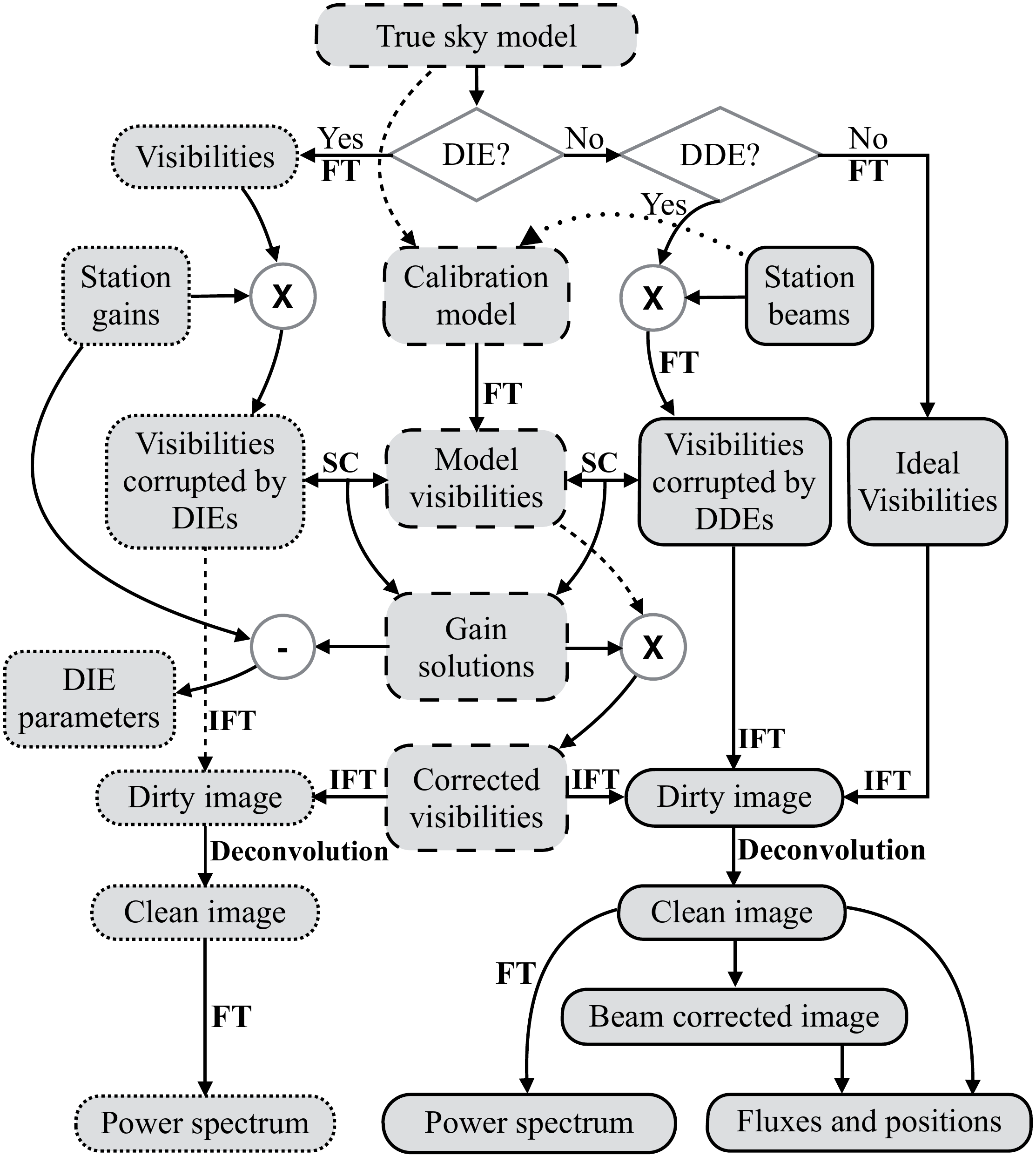}
\caption{Block diagram of the pipeline of the simulations of extragalactic foreground. Blocks with solid and dotted borders represent simulations with DD and DI errors respectively; blocks with dashed borders represent steps performed for both simulations, but separately. Arrows with dashed line-styles have been used to avoid intersection between arrows. FT, IFT and SC stand for Fourier transform, inverse Fourier transform and self-calibration respectively.}
\label{f:ppipe}
\end{figure}
A block diagram of the pipeline for simulating extragalactic point sources is shown in Fig. \ref{f:ppipe}. We start from a given model of the sky (described in the specific sections) and predict the visibilities that LOFAR would produce in the presence of certain DI and DD (beam) errors. Simulations with the two systematic errors are done separately, although some steps are common to both of them.

DI errors are introduced in accordance with the formulation described in section \ref{s:die}.
After prediction, the visibilities corrupted by the DIEs are self-calibrated using the same sky model that was used to predict.
Then, the gains determined by selfcal are compared with the input gains to calculate the error parameters defined by equations \ref{eq:diep}a-g.
Additionally, the solved gains are applied to the model visibilities to produce corrected visibilities.
All processes up to this point are performed using the standard LOFAR calibration and simulation software, Black Board Selfcal ({\tt BBS}; \citealt{pa09}).
We image both the corrupted and the corrected visibilities using {\tt CASA} and produce 2D PS from the images through the procedure described in section \ref{s:ps2d}.

\begin{table}
\begin{minipage}{\textwidth}
\caption{Observational setup for simulations of extragalactic sources:}
\label{t:setup1}
\begin{tabular}{@{}lllr@{}}
\hline
\hline
Number of LOFAR HBA stations used, $N$ & 59 \\
Number of baselines, $N(N-1)/2$ & 1711 \\
Number of spectral subbands & 1 \\
Number of channels in the subband & 1 \\
Central frequency of the channel & 150 MHz \\
Width of the channel, i.e. frequency resolution & 0.19 MHz \\
Total observation time & 8 h \\
Integration time, i.e. time resolution & 10 s \\
Number of timeslots & 2874 \\
Number of visibilities & 5090520 \\
Baseline cut ($u_{\rm min}\sim u_{\rm max}$) for imaging & 0.06 -- 20 km \\
Baseline cut for PS estimation & 0.06 -- 1 km \\
Angular resolution (PSF) of the images, $\alpha\lambda/u_{\rm max}$ & $\sim 0.34$ arcmin \\
Physical width of the HBA stations, $D$ & 30 m \\
FWHM of station primary beams, $\alpha\lambda/D$ & $\sim 3.78$ deg \\
Field of view, $\pi(\rm FWHM/2)^2$ & 11.2 deg$^2$ \\
\hline
\end{tabular}
\end{minipage}
\end{table}

DD errors are introduced by multiplying every point source in the model with the relevant station beam at the position of the source at every timeslot.
Fourier transform of the beam attenuated sky yields the visibilities corrupted by DDEs.
We carry out two different simulations with these dataset: one to measure effects of DD errors, and another to quantify the errors in calibration due to incomplete calibration sky model.
The latter could be done meaningfully without introducing systematic errors at all, but we did it this way to make it more realistic.

To quantify the effects of DD errors, first, we correct the corrupted visibilities for the beam at the phase centre which, in reality, normalizes the DDEs with respect to the phase centre so that only the differential nominal beam effects remain (this step is not shown in Fig. \ref{f:ppipe}).
Then, we image both the corrupted and uncorrupted (ideal) visibilities and produce 2D PS from the images.
Furthermore, we extract the fluxes and positions of the brightest point sources in the corrupted and uncorrupted images using PyBDSM\footnote{\url{http://tinyurl.com/PyBDSM-doc}} and compare them.
Finally, we correct the visibilities for the differential beam and produce images from them using {\tt AWImager}.
Fluxes of the beam-corrected images are compared with the uncorrected fluxes to quantify the quality of the correction.

To determine calibration errors due to an incomplete sky model, we calibrate the corrupted visibilities using different incomplete sky models.
As the same DDEs are included during both prediction and calibration, the remaining errors will be only due to the incompleteness of the models.
The deviation of the different corrected visibilities from the corrupted visibilities is demonstrated through PS.

\begin{figure*}
\begin{minipage}[b]{0.4\linewidth}
\centering
\includegraphics[width=\textwidth]{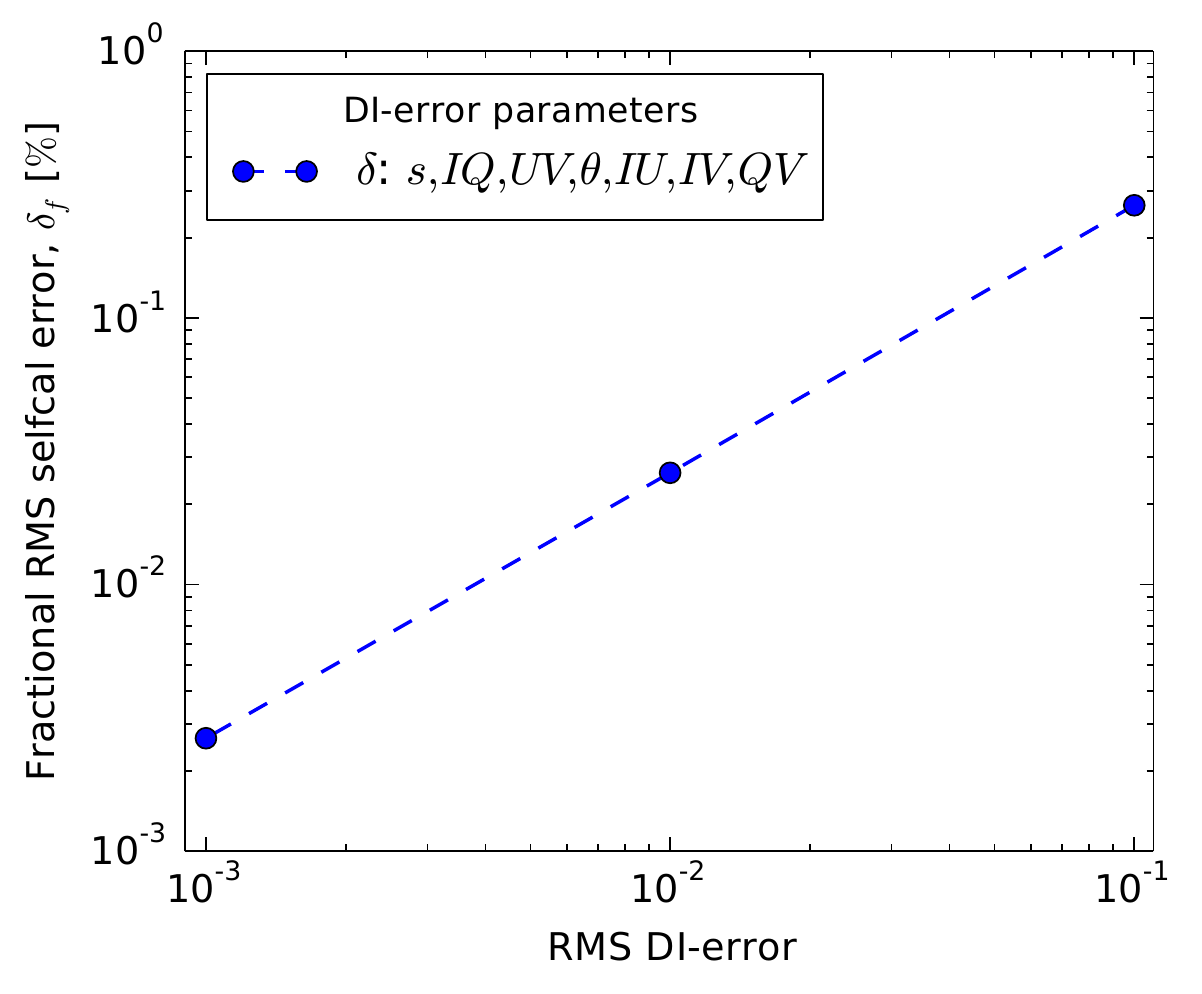}
\end{minipage}
\begin{minipage}[b]{0.4\linewidth}
\centering
\includegraphics[width=\textwidth]{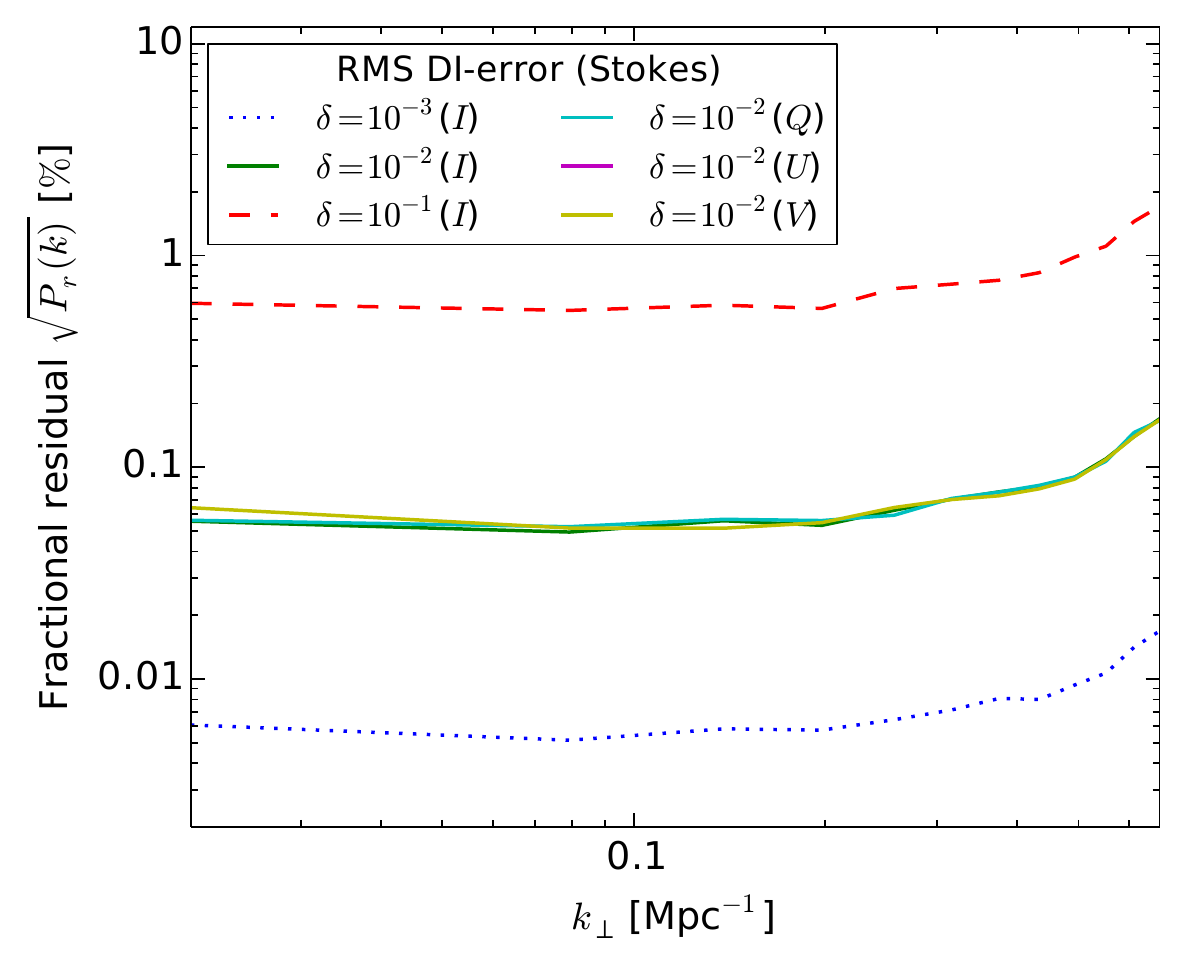}
\end{minipage}
\caption{\textit{Left}: Fractional error on the 7 DIE parameters defined in Eq. \ref{eq:diep} for a single baseline as a percentage of the input rms DI-errors.
The $\Delta g$ and $\Delta\epsilon$ used to calculate these parameters are the differences between the components of the input Jones matrix and the Jones matrix calculated by self-calibration.
\textit{Right}: Square-root of the fractional residual power spectra, which is equivalent to the rms of the image, for different rms DI-errors and Stokes parameters. See section \ref{s:sdie} for details.}
\label{f:die}
\end{figure*}

\subsection{Direction independent errors} \label{s:sdie}
To show the effects of DI errors and test their correction strategy, we ignore the DDEs and introduce DIEs for every station and timeslot as $G$-Jones matrices.
Both gain ($g$) and feed ($\epsilon$) error terms of $\mathbf{G}$ are modelled as complex numbers that are random at every time-step drawn from a Gaussian distribution with zero mean and a certain standard deviation (rms).
Then, we create a sky model containing 25 sources of 5 Jy Stokes $I$ flux ($Q,U,V=0$) in a $5\times 5$ uniform grid of $1^\circ$ separation, predict the DIE-corrupted visibilities for all baselines of LOFAR and perform all the other steps described in the previous section and shown in Fig. \ref{f:ppipe} (see the blocks with dotted and dashed borders).
The rms of the introduced errors is the same for every term of the $G$-Jones of every station and we repeat this experiment thrice for three different rms DI-errors: $10^{-3}$, $0.01$ and $0.1$.
Note that, as the calibration was done with a perfect sky model, the errors will be due only to the calibration process itself.

We analyse the results using two parameters: fractional rms selfcal error ($\delta_f$) and square-root of the residual power spectrum ($\sqrt{P(k)}$) which, in effect, gives the rms of the images at different spatial frequencies.
To determine $\delta_f$, we calculate $\Delta\mathbf{G}^g$ and $\Delta\mathbf{G}^f$ (see Eq. \ref{eq:delG}) by differencing the model gains and the solved gains for two stations, and then, calculate the DIE-parameters ($\delta$) for the baseline created by those stations (equations \ref{eq:diep}).
We did not plug in the values of $\Delta\mathbf{G}^g$ and $\Delta\mathbf{G}^f$ directly in Eq. \ref{eq:delG} to calculate $\delta$, but created an error DI-Mueller matrix from the $\Delta\mathbf{G}$ matrices of two stations following Eq. \ref{eq:m2j} and extracted the 7 relevant parameters from it.
$\delta_f$ for a given $\delta$ is the rms of the $\delta$ as a percentage of the input rms DI-error.
The seven $\delta_f$ are plotted as a function of the input rms DI-errors on the left panel of Fig. \ref{f:die}.
We see that fractional selfcal errors increase linearly with rms DI-errors, and for an rms DI-error of $10^{-3}$, which is not unrealistic, the error on these parameters is less than 0.002\%.

For calculating residual $P(k)$, we subtract the corrected Stokes images from the corrupted ones and measure the PS of the residuals.
As we did not subtract any source from the corrected visibilities, if the calibration error is low the difference between the corrected and the corrupted visibilities should also be low.
$P_r(k)$ is the PS of a residual image as a percentage of the PS of a corrupted image.
$\sqrt{P_r(k)}$ of the different Stokes images for the three simulations are plotted on the right panel of Fig. \ref{f:die} which clearly shows that the calibration errors propagated to the PS are negligible as expected in the absence of additive noise.
For an rms DI-error of $10^{-3}$, errors on $\sqrt{P(k)}$ or, equivalently, on the rms of the image is less than 0.005\%.
Furthermore, by comparing the Stokes $I$, $Q+iU$ and $V$ power spectra for an rms DI-error of 0.01, we see that the errors on different Stokes parameters are the same, as expected.
This simulation shows that self-calibration can correct for the DI-errors to a very high accuracy if we have a sufficiently accurate model of the sky.

\subsection{Direction dependent errors}
To show the effects of DD errors on point sources and to test one of their correction strategies, we ignore the DIEs, introduce DDEs as station beams and carry out the steps outlined in Fig. \ref{f:ppipe} (see the blocks with solid and dashed borders).
As mentioned before, we implemented two different simulations with the DDE-corrupted dataset; the purpose of the first one is to show the effects of DD-errors on the Stokes parameters and this has been done for two different sky models, some information about which are listed in table \ref{t:sky}.

\subsubsection{Test with a mock sky} \label{s:mock}
To show the general trend of the effects of DD-errors, we make a mock sky model comprising 225 unpolarized point sources arranged in a $15\times 15$ uniform grid of $0.66^\circ$ separation centred on the position of 3C196\footnote{A quasar situated at $z\sim 0.871$ with a flux density of 74.3 Jy at 174 MHz.} and simulate an 8-hour, 150-MHz observation of LOFAR, taking into account the beams described in section \ref{s:dde}.
The source at the centre of the grid is given a flux density of 100 Jy, while each of the other sources have a flux density of 0.4 Jy.
The central source has been made exceptionally bright (analogous to the 3C196 field) to be able to check the consequence of calibrating an otherwise dim sky with a very bright point source which will be described in section \ref{s:isky}.

\begin{figure*}
\centering
\includegraphics[width=\textwidth]{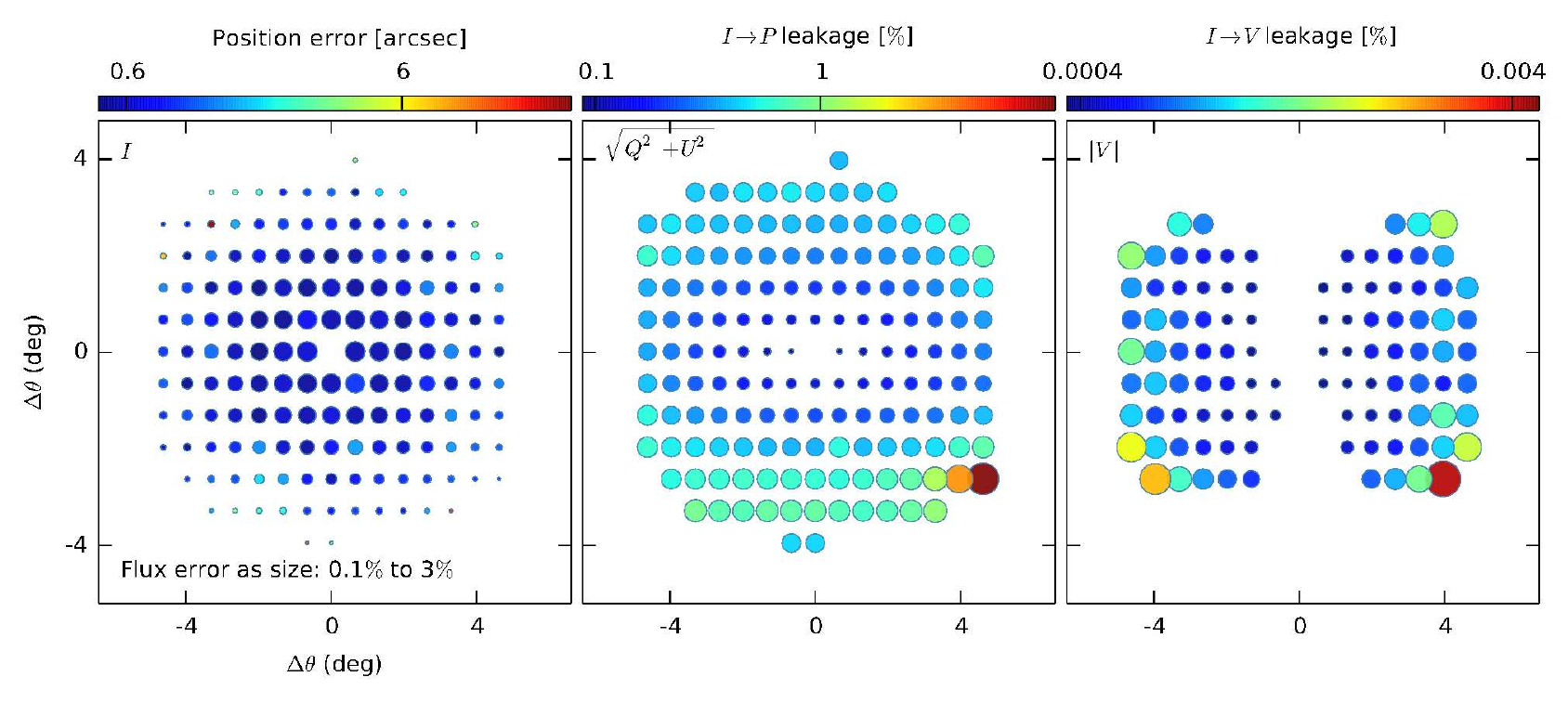}
\caption{\textit{Left}: Distribution of 103 sources from the 225 sources arranged in a 15$\times$15 uniform grid within a $10^\circ$ field of view. Flux (bubble size) and position (colour) errors due to calibration with only the prominent central source are shown. \textit{Middle}: Same distribution with corresponding fluxes leaked from Stokes I to linear polarization as a percentage of Stokes I flux (size and colour). \textit{Right}: Same as the middle figure except that it is for the leakages to circular polarization which is much lower.}
\label{f:mock}
\end{figure*}

As the sources were completely unpolarized, the Stokes $Q,U,V$ images created from this dataset contain only the flux leaked from Stokes $I$, i.e. instrumentally polarized sources.
These sources are shown on the middle and right panels of Fig. \ref{f:mock}.
Each bubble in the plots represent an instrumentally polarized point source and the size and colour of the bubble represent the flux of the source as a percentage of its Stokes $I$ flux.
The figures show that leakages to both linear and circular polarizations increase as we go out from the centre of the field.
As for the levels of leakage, within the central 4 degrees, i.e. within the first null of the primary beam at 150 MHz, linear polarization leakage ($I\rightarrow P$) is around 0.5\%, and circular polarization leakage ($I\rightarrow V$) is less than 0.003\%.
Instrumental polarization of the central bright source (not shown in the figure) is very low, because before imaging the visibilities corrupted by the DDEs were corrected for the element beam ($\mathbf{E}_e$ of Eq. \ref{eq:dbeam}) at the phase centre, thereby making the leakage terms very close to zero at that point.
In physical terms this means that the projection of the beams on the sky had been made perfectly orthogonal at the phase centre.
What is left after this centre-correction is the effect of the differential beam (e.g. Fig. \ref{f:mueller}a).
There is an anomaly in the south-east corner of the middle and the right panels of Fig. \ref{f:mock} which can be attributed to the errors in extracting fluxes of very dim sources situated near the null of the primary beam.

\begin{table*}
\centering
\begin{minipage}{\linewidth}
\caption{Sky models used for the different simulations of extragalactic foreground with DDEs.}
\label{t:sky}
\begin{tabular}{@{}lllllllllr@{}}
\hline
Field & Phase centre & Phase centre & FoV & Catalogue & Number of & Maximum & Minimum & Total\footnote{All flux densities shown here are at 150 MHz.} & Spectral index \\
 & (Equatorial J2000) & (Galactic) & (deg) & & sources & (Jy) & (Jy) & (Jy) &  \\
\hline
Mock & $\alpha\sim 8^h13'36''$, $\delta\sim 48^\circ 13'0''$ & $l\sim 171^\circ$, $b\sim 33^\circ$ & 10 &  & 225 & 100 & 0.4 & 189.6 & -0.75 \\
3C196 & $\alpha\sim 8^h13'36''$, $\delta\sim 48^\circ 13'0''$ & $l\sim 171^\circ$, $b\sim 33^\circ$ & 10 & FIRST\footnote{The \textbf{F}aint \textbf{I}mages of the \textbf{R}adio \textbf{S}ky at \textbf{T}wenty-cm survey, produced by NRAO VLA at 1365 and 1435 MHz and $5''$ resolution; noise $\sim 0.15$ mJy.} & 4567 & 83 & 0.027 & 796.64 & -0.75  \\
\hline
\end{tabular}
\end{minipage}
\end{table*}

These results are consistent with the beam model described in section \ref{s:dde}.
For example, we can understand both the trend and the level of linear leakage seen in Fig. \ref{f:mock} by comparing it to the $M_{21}$ and $M_{31}$ components of the instrumental Mueller matrix shown Fig. \ref{f:mueller}a, or to the spatio-temporal profiles of the leakages shown in Fig. \ref{f:mueller}b.
We expect to see leakage at this level also in the realistic simulations and this expectation will be put to the test in the next section where we describe the simulation of one of the LOFAR-EoR target fields.
\begin{figure*}
\centering
\includegraphics[width=0.9\linewidth]{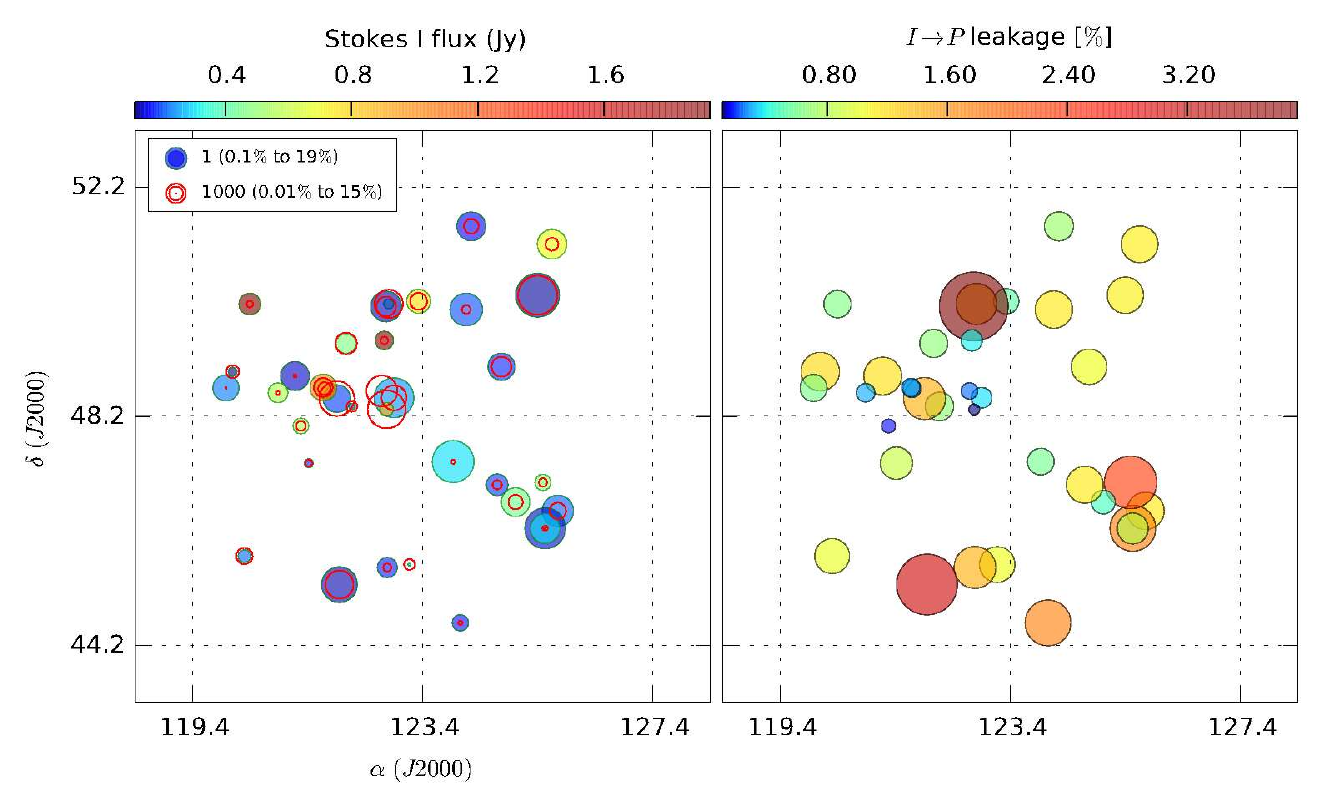}
\caption{\emph{Left}: Distribution of the brightest 33 sources ($I>100$ mJy) in the 3C196 field with their corresponding Stokes I fluxes (colour) and flux errors (bubble size) due to calibration with different number of sources (numbers in the legend) in the sky model. The percentages in the legend refer to the minimum and maximum flux errors. Note that after calibration with 1000 sources errors for most of the sources decrease. \emph{Right}: Same distribution with the corresponding linear polarization leakages as a percentage of Stokes $I$ flux. Both colour and size of the bubbles represent fractional leakage.}
\label{f:3c}
\end{figure*}
\subsubsection{3C196 field}
The 3C196 field (centred on the bright quasar, 3C196; \citealt{be10}) is well-suited for EoR observations because the presence of a bright and almost unresolved source at its centre allows very accurate direction independent calibration, and it is situated in one of the colder regions of the Galactic halo.
To make an unpolarized sky model for simulating this field, we extract Stokes $I$ fluxes and positions of the sources brighter than 25 mJy within a radius of $5^\circ$ around 3C196 from the FIRST survey catalogue (see table \ref{t:sky}) and extrapolate the fluxes to that of 150 MHz using a spectral index of -0.75 which is typical for the radio sources at these frequencies.
The eponymous source, 3C196, has been taken out of this model, and a 4-component improved model of the source made from LOFAR data by V. N. Pandey has been inserted in its place.

Linear leakage of the brightest 33 sources (Stokes $I>100$ mJy) is shown on the right panel of Fig. \ref{f:3c}.
Both colour and size of the bubbles in the figure represent the percentage of leakage.
Extraction of fluxes and positions of the sources in this case is not as precise as that of the gridded sky model as here sources are much more closely spaced; thus some errors in this scatter plot originate from the source extraction process.
Nevertheless, the figure, as a whole, is quite informative; we see that linear leakage can be as high as $4\%$, but for most of the sources it is less than $2\%$ and for the sources very close to the phase centre only less than a percent leak, as expected.
The sources with the highest leakages (the three reddest bubbles) are very dim in Stokes $I$ which can be seen by comparing these three bubbles with the corresponding bubbles on the left panel where the Stokes $I$ fluxes are shown as colour of the bubbles.
These leakages might not be real, but a consequence of errors in the source extraction process.
The leakage from 3C196 itself is very low and hence is not shown here.

\begin{figure*}
\centering
\begin{minipage}[b]{0.4\linewidth}
\includegraphics[width=\textwidth]{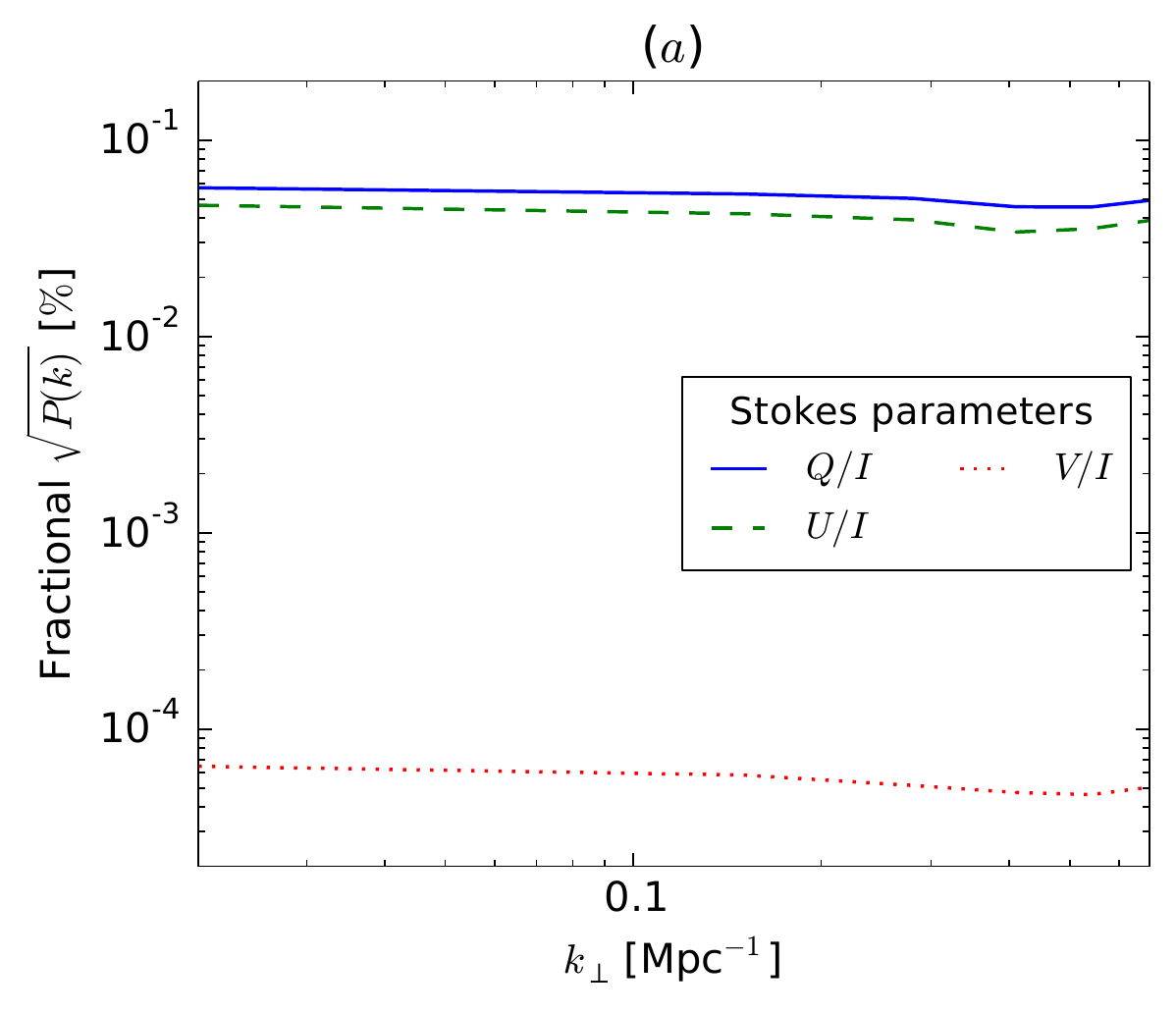}
\end{minipage}
\begin{minipage}[b]{0.38\linewidth}
\includegraphics[width=\textwidth]{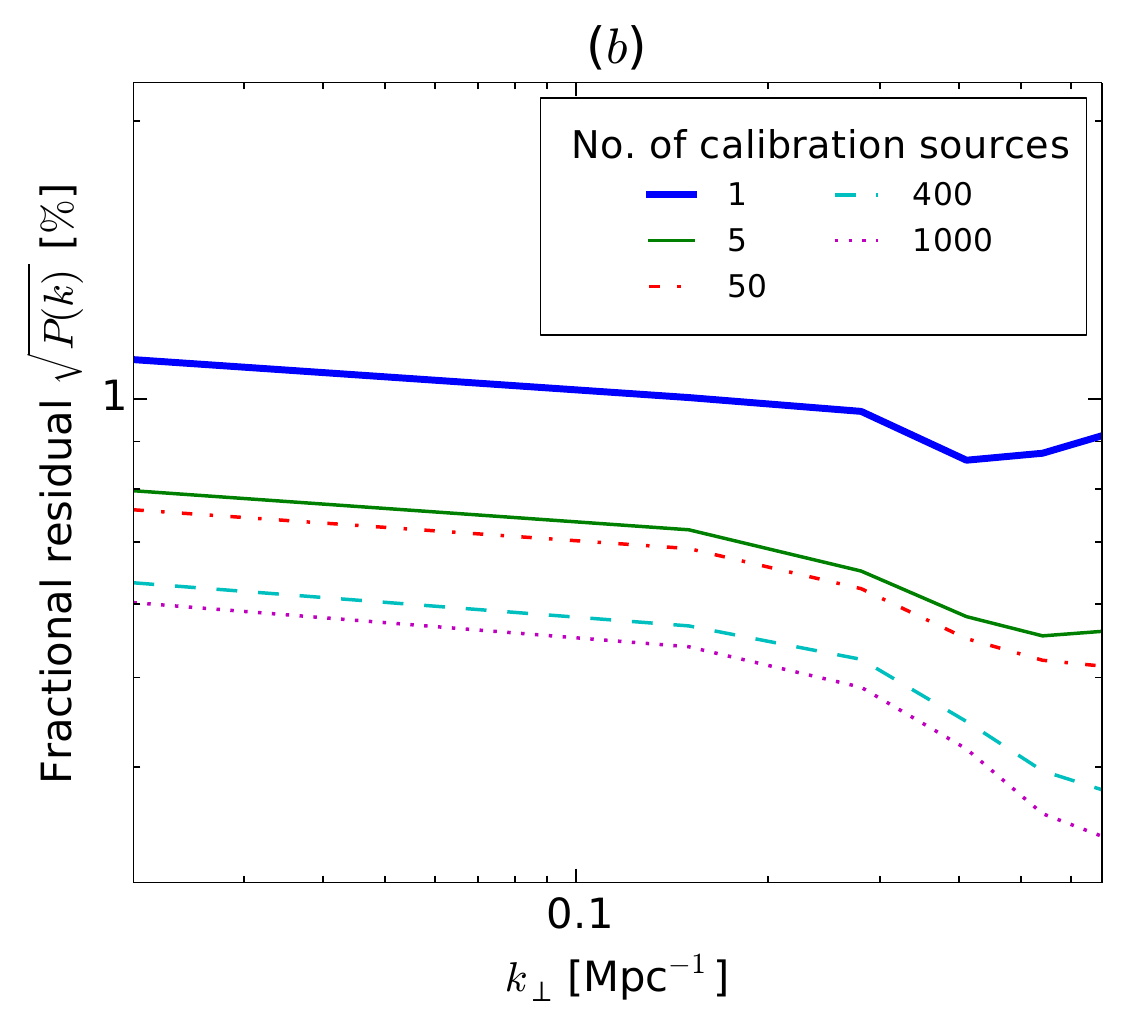}
\end{minipage}
\caption{\emph{a}: Square-root of the power spectra of the $Q,U,V$ leakages as a percentage of the Stokes $I$ power spectrum within the central 10 degrees of the 3C196 field.
\emph{b}: Residual (after subtracting calibrated data from the uncalibrated ones) power spectra of Stokes $I$ as a percentage of the uncalibrated Stokes $I$ PS of the same field.
The different cases are for calibration with different number of sources in the sky model.
These residuals correspond to calibration errors due to incomplete sky model.}
\label{f:pspoints}
\end{figure*}

The overall level of the leakage can be better understood from the fractional (as a percentage of Stokes $I$ PS) power spectra of Stokes $Q,U,V$ shown in Fig. \ref{f:pspoints}a.
The PS of $Q/I$ and $U/I$ tell us that the rms of the linear leakage is $0.05\sim 0.06\%$ of the rms of the Stokes $I$ image.
On the other hand, rms of circular leakage is almost 4 orders of magnitude lower.
Leakage from linear polarization to Stokes $I$, which is relevant for the EoR experiments, will be similar to the $I\rightarrow P$ leakage shown in this simulation, as evident from a comparison of the first and second panels from the top of Fig. \ref{f:mueller}b.
However, compact radio sources are usually unpolarized or very weakly polarized and hence the leakage from polarized point sources into Stokes $I$ is very low and even that leakage can be removed by direction dependent calibration (e.g. {\tt SAGECal}; \citealt{ka11,ka13}) and/or AW-projection \citep{ta}.

\subsubsection{Correcting polarization leakage of point sources}
There are several strategies for correcting beam-related DDEs which are classified broadly into two categories: image-plane and Fourier-plane corrections.
Here, we test one of the Fourier-plane strategies called AW-projection (see section \ref{s:aw}), a particular version of which is implemented by {\tt AWImager} for the LOFAR AW-terms.

To do a simple test, we create a dataset from a sky model consisting of 36 unpolarized 10 Jy sources in a $6\times 6$ uniform grid of 0.5 degree separation so that all sources are within the FWHM of the primary beam, and then try to correct the Stokes $I$ fluxes and remove the leakages using {\tt AWImager}.
As mentioned in section \ref{s:aw}, LOFAR $\mathcal{A}$-terms are separated into two parts by {\tt AWImager}: the slowly varying (in time) element beam ($\mathbf{E}_e$), and the fast-varying array factor.
We assume $\mathbf{E}_e$ to be constant within 12 minutes, and the array factor to be constant within 5 minutes.

\begin{figure}
\centering
\includegraphics[width=0.8\linewidth]{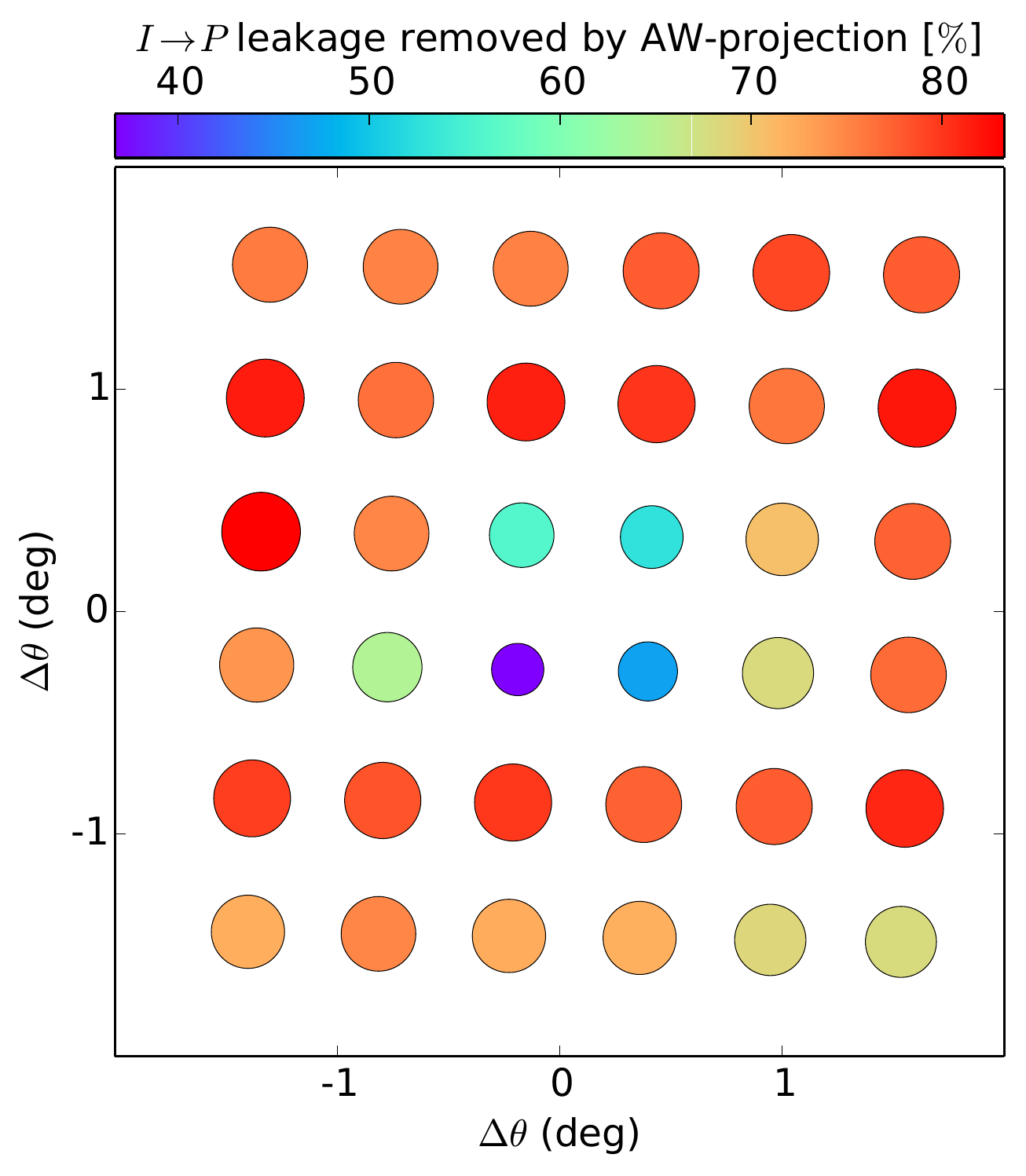}
\caption{Linear polarization leakage removed by {\tt AWImager} as a percentage of the leakage; both size and colour of the bubbles represent the same quantity. All sources in this simulated dataset had a Stokes $I$ flux of 10 Jy and their linear leakages were around 0.1 mJy. We see that up to 80\% leakage could be removed using {\tt AWImager} in this case.}
\label{f:aw}
\end{figure}

The result is shown in Fig. \ref{f:aw}; both color and size of the bubbles represent percentage of leakage removed by {\tt AWImager}.
It seems that up to 80\% of the leakage can be removed.
The performance appears to be worse near the centre of the field than further away which is counter-intuitive, but the leakage is already very low near the centre and the bad performance could be due to the inefficiency of both {\tt AWImager} and the flux extraction software in dealing with faint sources.
We should be careful to draw any final conclusions on the effectiveness of {\tt AWImager} in removing leakages from our data as the software is still under construction and we are not aware of any test of leakage removal done on a realistic dataset.

\subsection{Selfcal errors due to incomplete sky model} \label{s:isky}
Incomplete sky models can lead to many problems in directionally independent self-calibrated data, among them generation of spurious source components, removal of real source components and the generation of ghost sources, a spurious source whose flux is proportional to the flux of an unmodelled source \citep{gr}.
We try to quantify the calibration errors due to incomplete sky models with different numbers of sources in the models for the 3C196 field.
Note that, as we included the beam during both prediction and calibration, its effect was taken out and we were left with only selfcal errors.
The calibration is performed using {\tt BBS} which is based on the matrix formalism described in section \ref{s:form}.

We make 5 different calibration sky models that contain roughly 10, 15, 30, 60 and 75\% of the total flux of the field; the models have 1, 5, 50, 400 and 1000 sources respectively.
After self-calibrating the field with each one of these models, we calculate the difference of fluxes of the sources between the calibrated and uncalibrated data, and also create corresponding residual PS.
The left panels of Fig. \ref{f:3c} show the flux errors on the sources that contribute to the largest errors in the field; the filled bubbles represent the errors after calibrating with only 10\% of the total flux, while the unfilled bubbles with red borders are for the case when 75\% flux is modelled.
According to the figure, errors go down significantly after improving the sky models.

In Fig. \ref{f:pspoints}b, we show the PS of the Stokes $I$ residual after subtracting the calibrated images from the uncalibtrated ones as a fraction of uncalibrated Stokes $I$ PS.
As there is an exceptionally bright source (the second brightest source is only 7.7 Jy) at the centre of the 3C196 field, rms of the residual is already low (1\% of the rms of the original image) after calibrating with only 3C196 which contains 10\% of the total flux of the field.
Errors go down significantly when we include 15\% of the total flux by adding another 4 sources in the model, but after that there is no rapid improvement.

\section{Simulation of Galactic foreground} \label{s:gal}
So far we have considered leakages from unpolarized point sources into Stokes $Q,U,V$ only, but, as mentioned before, our interest lies in the opposite case, i.e. leakage from polarization to total intensity.
Compact radio sources are very weakly polarized and most of the point sources seen in polarization maps can be attributed to instrumental polarization and leakage.
As at frequencies of tens to hundreds of MHz the polarized sky is dominated by Galactic diffuse synchrotron emission, we take real data of the 3C196 field observed by LOFAR, and create the simulated dataset using it as a sky model following the pipeline described in the next section.
In these simulations, except for the one represented by Fig. \ref{f:ps3dc}h, our Stokes $I$ data contain only the noise leaked from Stokes $Q,U$.
We do not add realistic noise to Stokes $I$ until the final test because that would make the quantification of the intrinsic instrumental polarization over the complete $k$-space difficult, as the expected level of leakage is lower than the system noise.
However, as a final test we add system noise to check the efficiency of a leakage removal technique.

\subsection{Simulation setup}
The general pipeline of the simulation of Galactic foreground is almost same as that of the extragalactic foreground (boxes with solid borders in Fig. \ref{f:ppipe}), but there are two major differences: here we simulate datasets for 161 spectral bands instead of just one, and examine the leakages from Stokes $Q,U$ to $I,V$ rather than that from $I$ to $Q,U,V$.
The former enables us to examine the frequency behaviour of the leakages through rotation measure synthesis and 3D power spectrum analysis, and the latter provides us with a realistic estimate of the amount of leakage into Stokes $I$ to be expected in the current LOFAR-EoR observations of the 3C196 field.

\begin{figure*}
\begin{minipage}[b]{0.25\linewidth}
\includegraphics[width=\textwidth]{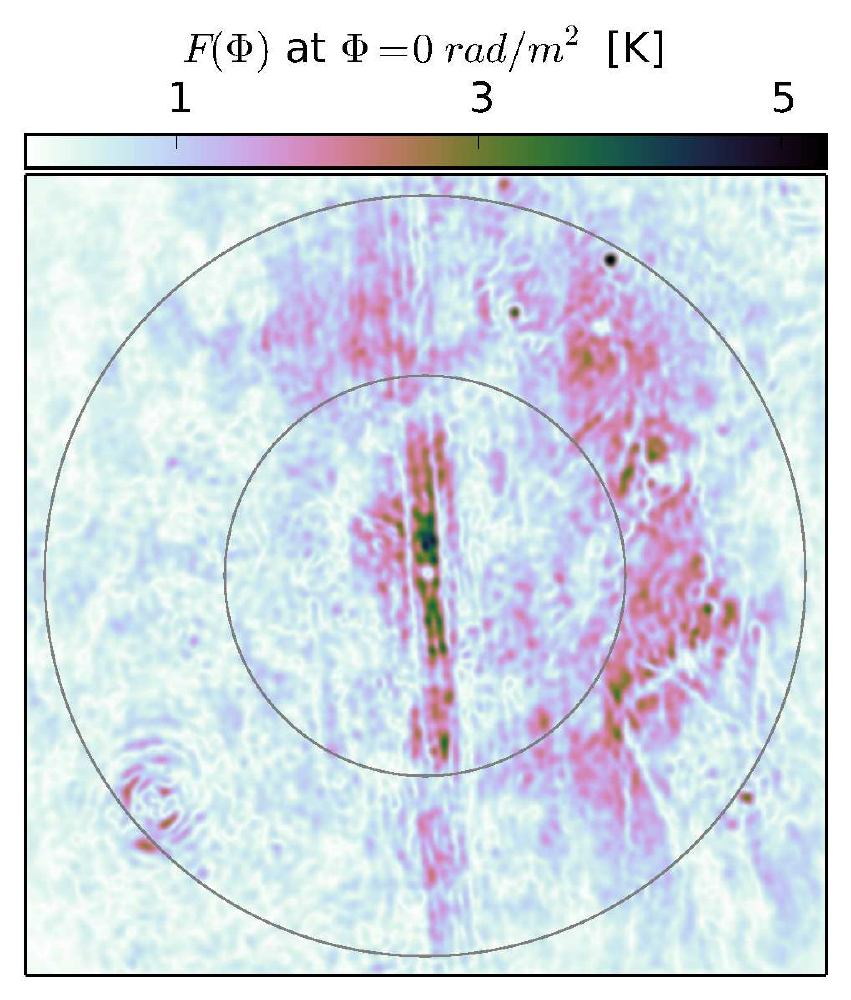}
\end{minipage}
\begin{minipage}[b]{0.25\linewidth}
\includegraphics[width=\textwidth]{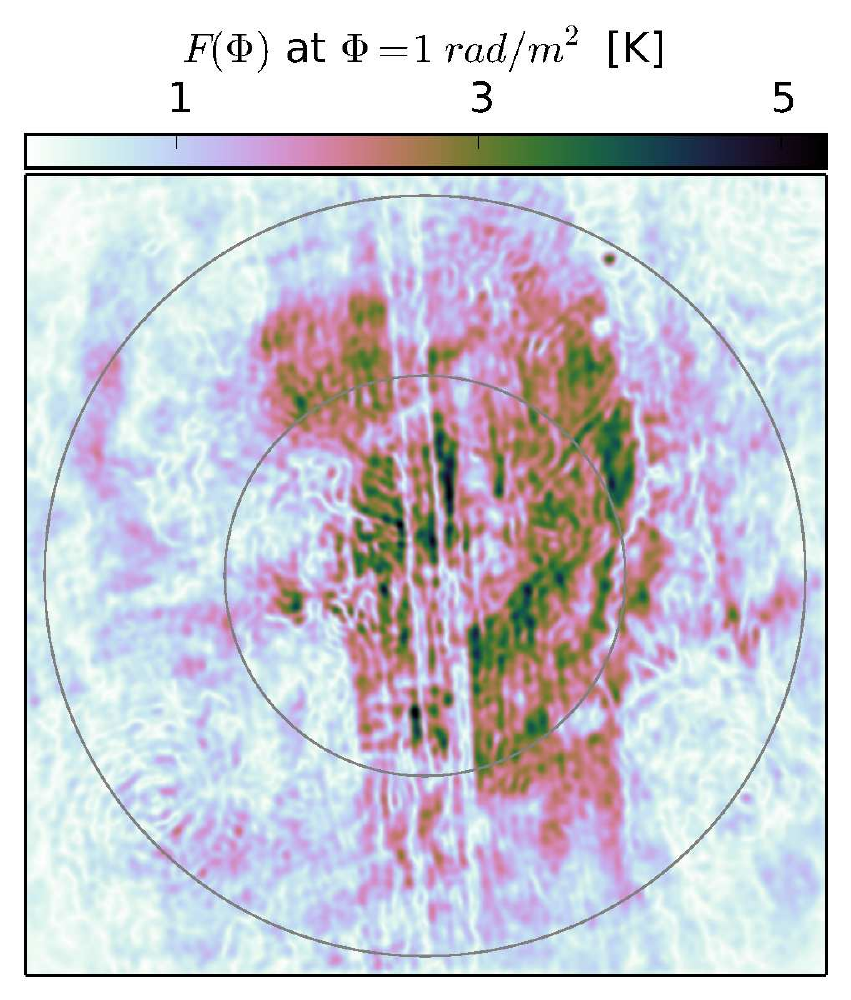}
\end{minipage}
\begin{minipage}[b]{0.25\linewidth}
\includegraphics[width=\textwidth]{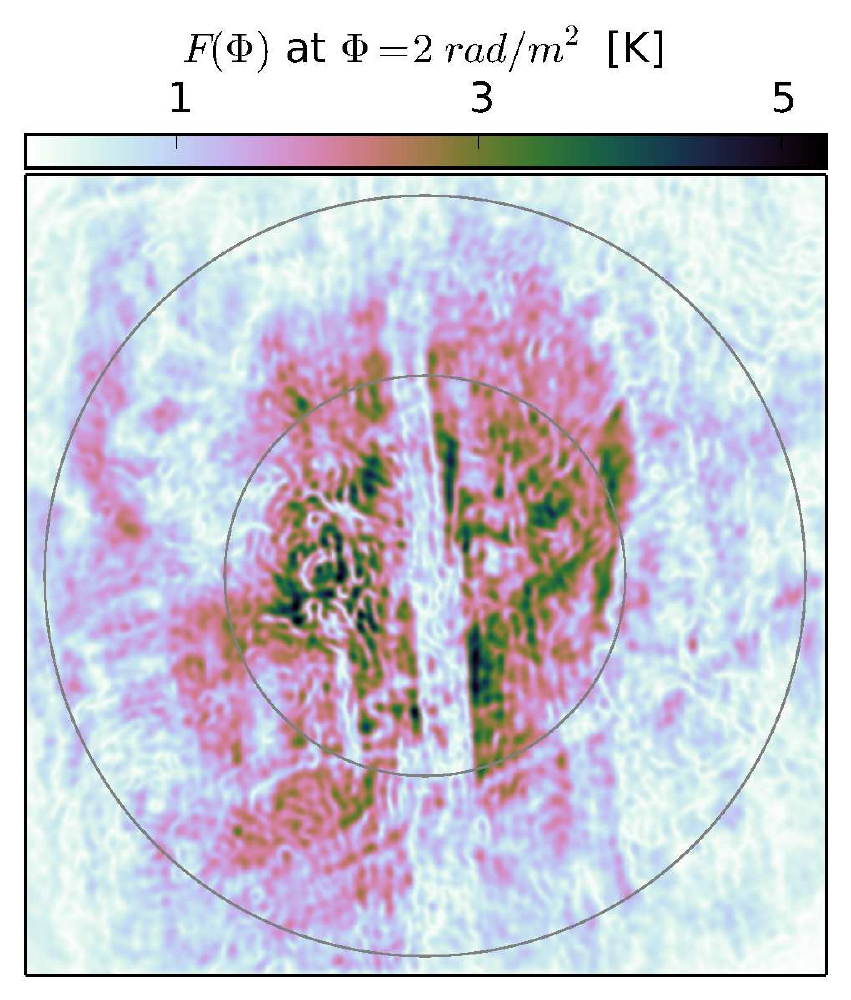}
\end{minipage}

\begin{minipage}[b]{0.25\linewidth}
\includegraphics[width=\textwidth]{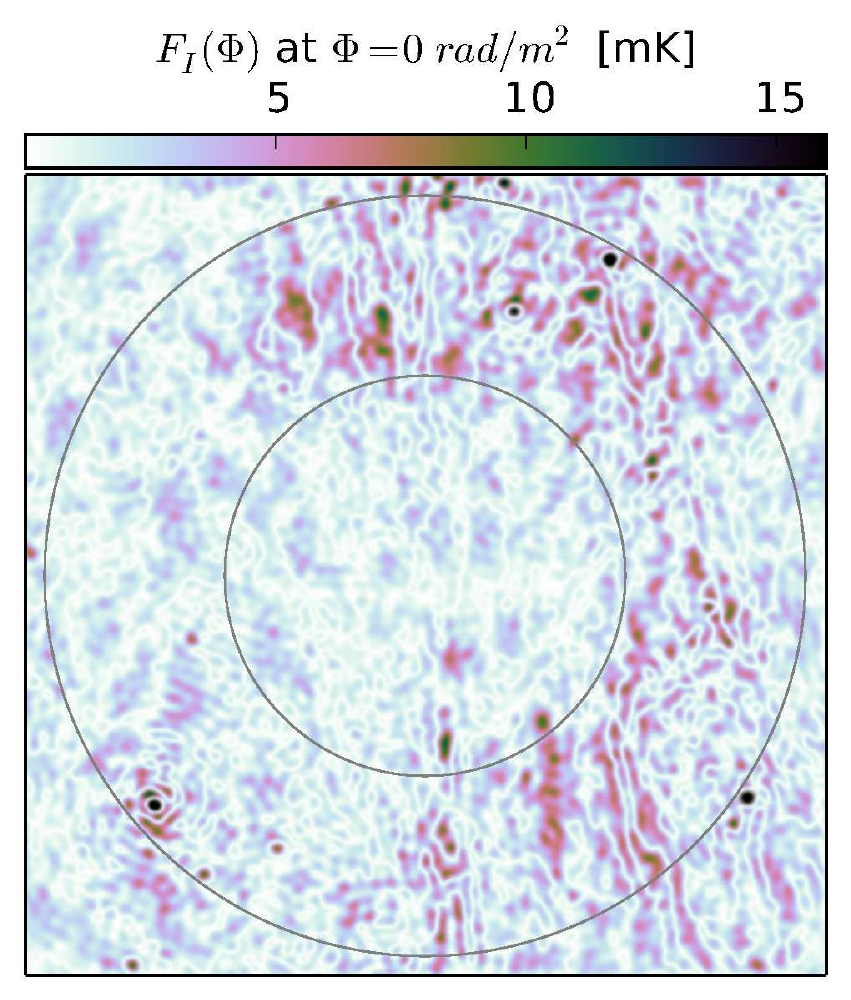}
\end{minipage}
\begin{minipage}[b]{0.25\linewidth}
\includegraphics[width=\textwidth]{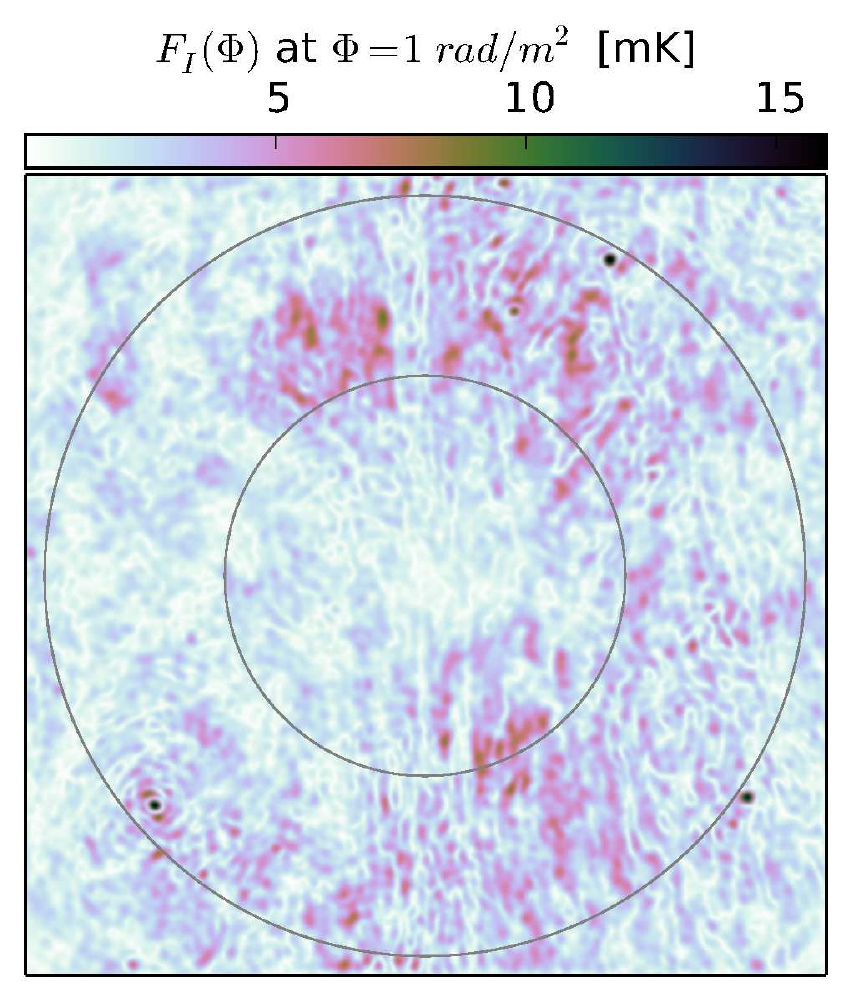}
\end{minipage}
\begin{minipage}[b]{0.25\linewidth}
\includegraphics[width=\textwidth]{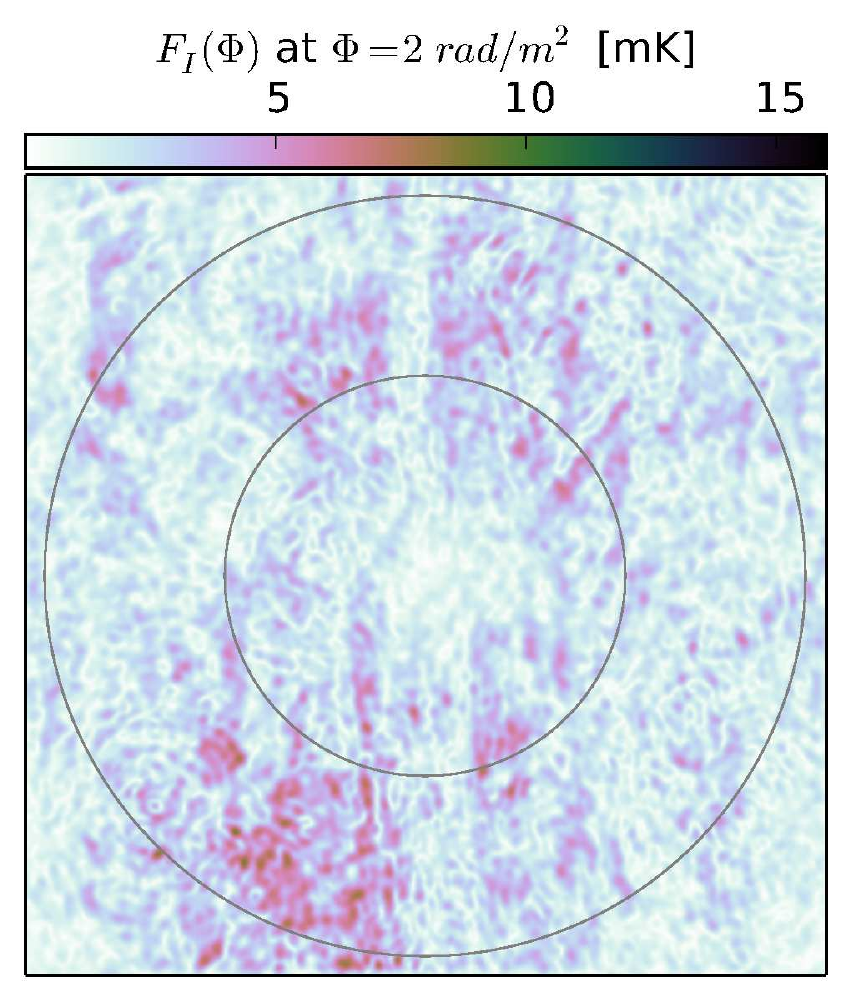}
\end{minipage}

\begin{minipage}[b]{0.25\linewidth}
\includegraphics[width=\textwidth]{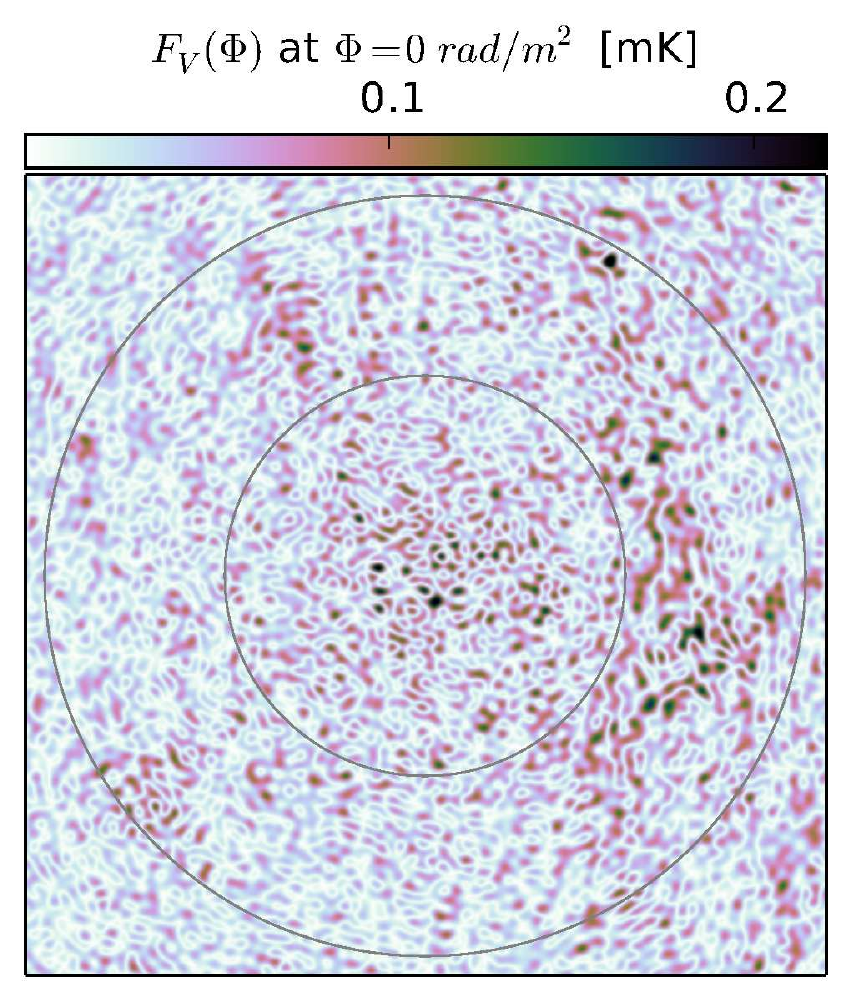}
\end{minipage}
\begin{minipage}[b]{0.25\linewidth}
\includegraphics[width=\textwidth]{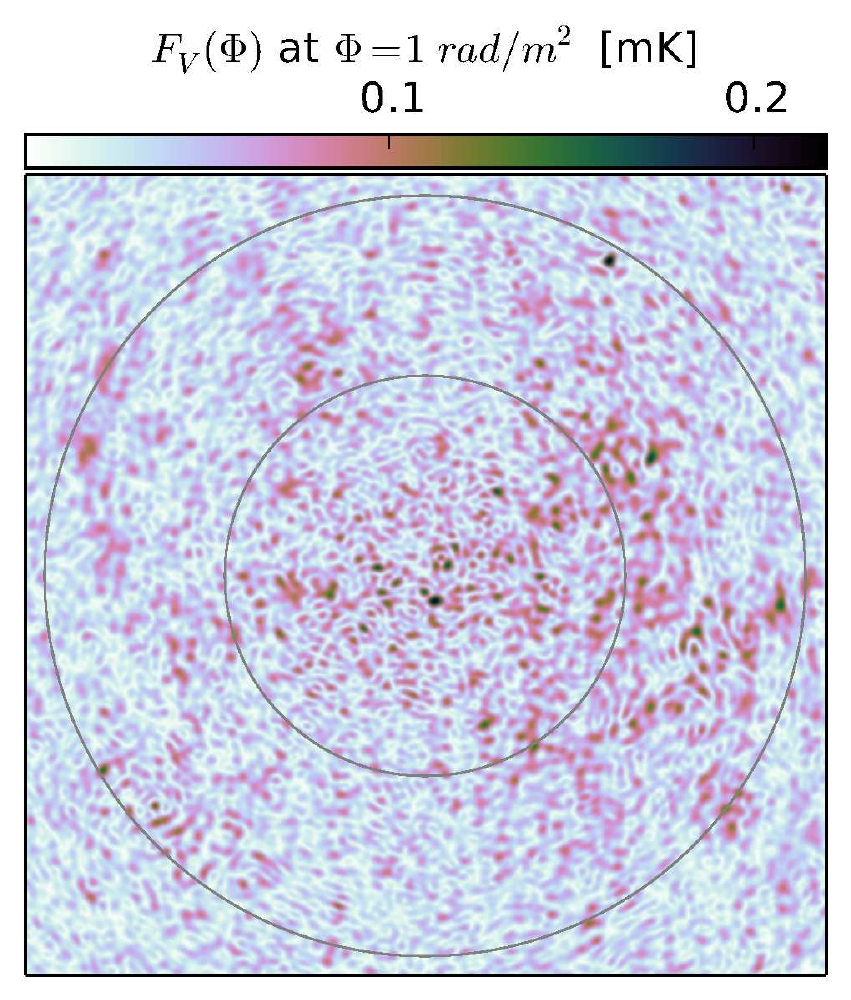}
\end{minipage}
\begin{minipage}[b]{0.25\linewidth}
\includegraphics[width=\textwidth]{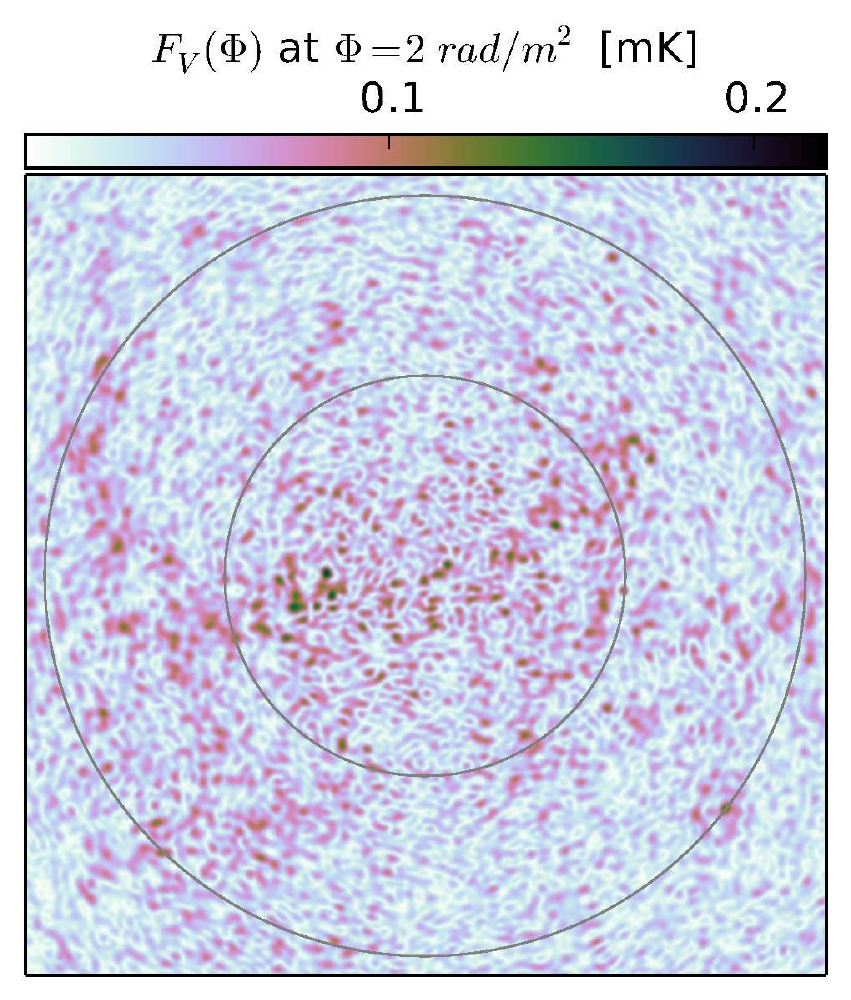}
\end{minipage}
\caption{Faraday dispersion images of Galactic diffuse polarized emission within the central 4 degrees of the 3C196 field (top) and their leakages to Stokes $I$ (middle) and $V$ (bottom) caused by the LOFAR differential beams at the Faraday depths of 0 (left), +1 (middle) and +2 (right) rad/m$^2$.
The diameters of the inner and the outer circles are $2^\circ$ and $3.8^\circ$ (FWHM of LOFAR station beam at 150 MHz) respectively.
The images have $480\times 480$ pixels of $0.5'$ with a PSF of $3'$.}
\label{f:rmimg}
\end{figure*}

To make the 3C196 polarization sky model, we took Stokes $Q,U$ images for 161 subbands spanning 32 MHz centred at 150 MHz that were produced from a single-night (8 hr) LOFAR observation using the standard LOFAR calibration and imaging pipeline \citep[e.g. see][]{ya13,je14}.
During the reduction process, DI-errors were removed and the data was also corrected for the element beam at the phase centre, thereby removing most of the instrumentally polarized point sources.
We removed the remaining point sources by just masking them with noise so that diffuse emission dominates the image.
The most significant systematic errors that still remains in these images are the ionospheric Faraday rotation and the differential beam.
A dataset with ionospheric correction implemented is not necessary for our case as we are not concerned with analysis of the real data here, but only with the fraction of leakage; thus, any reasonable input model would serve our purpose.
Also note that we are applying the `model' differential beam to an image that already has the `true' differential beam in it.
This cannot be avoided as direction dependent calibration or differential beam correction are yet to be done in this observing window, but DD-correction would not bring any dramatic change in the final results that we want to produce, as the polarization maps are dominated by diffuse emission and differential beam is only an 1\% effect.

\begin{table}
\begin{minipage}{\textwidth}
\caption{Setup of the simulation of Galactic foreground.}
\label{t:setup2}
\begin{tabular}{@{}lllr@{}}
\hline
\hline
Baselines used for simulated observation & up to 3 $k\lambda$ \\
Number of spectral subbands & 161 \\
Number of channels in each subband & 1 \\
Width of the channels / frequency resolution, $\delta\nu$ & 0.19 MHz \\
Central frequency of the observing band & 150 MHz \\
Total bandwidth, $\Delta\nu$ & 32 MHz \\
Total observation time & 8 hr \\
Integration time / time resolution & 10 s \\
Baseline cut for imaging and PS estimation & 30 -- 800 $\lambda$ \\
Angular resolution (PSF) of the images & 4.3 arcmin \\
Number of pixels in the images & $480\times 480$ \\
Size of each pixel & 0.5 arcmin \\
Maximum detectable Faraday depth, $\sqrt{3}/\delta\lambda^2$ & 160 rad/m$^2$ \\
Largest resolvable structure in Faraday depth, $\pi/\lambda_{min}^2$ & 0.96 rad/m$^2$ \\
Resolution in Faraday depth space, $2\sqrt{3}/\Delta\lambda^2$ & 1 rad/m$^2$ \\
Minimum and maximum $k_\perp$ [Mpc$^{-1}$] & 0.02 -- 0.53 \\
Minimum and maximum $k_\parallel$ [Mpc$^{-1}$] & 0.011 -- 1.85 \\
\hline
\end{tabular}
\end{minipage}
\end{table}

Stokes $I$ and $V$ in the model images were put to zero so that after applying the beam, they contain only leakages from $Q,U$. The following steps were performed to produce the final results.
\begin{enumerate}
\item DDE-corrupted visibilities at different frequencies are simulated using {\tt AWImager}, as a prediction using {\tt BBS} would currently take too much time.
Here {\tt AWImager}, in effect, carries out the forward transform of the major cycle and stops.
\item Images from the simulated visibilities are produced using {\tt CASA}.
Different parameters of the input model images and the final {\tt CASA} images were kept the same; for details see table \ref{t:setup2}.
\item We make 4 image-cubes by combining the images for 4 Stokes parameters and also convert the fluxes in Jy to intensities in temperature following Eq. \ref{eq:conv}.
\item To analyse Faraday structure of the leakage of polarized emission, RM-synthesis is performed on the cubes according to the formalism of section \ref{s:rm} resulting in 3 `dirty' (without deconvolving the RMSF) RM-cubes: $F(\Phi)$, $F_I(\Phi)$ and $F_V(\Phi)$.\footnote{The RM-synthesis code of Michiel Brentjens was used for this purpose.}
\item Cylindrical and spherical 3D power spectra for all Stokes parameters are calculated from the image-cubes according to the formalism described in section \ref{s:ps3d}.
\end{enumerate}

\subsection{Results}
We first show the results of RM-synthesis of both polarization and leakage, but with a focus on the leakages, and then present the 3D power spectra produced from the image-cubes.

\subsubsection{RM synthesis}

\begin{figure}
\includegraphics[width=\linewidth]{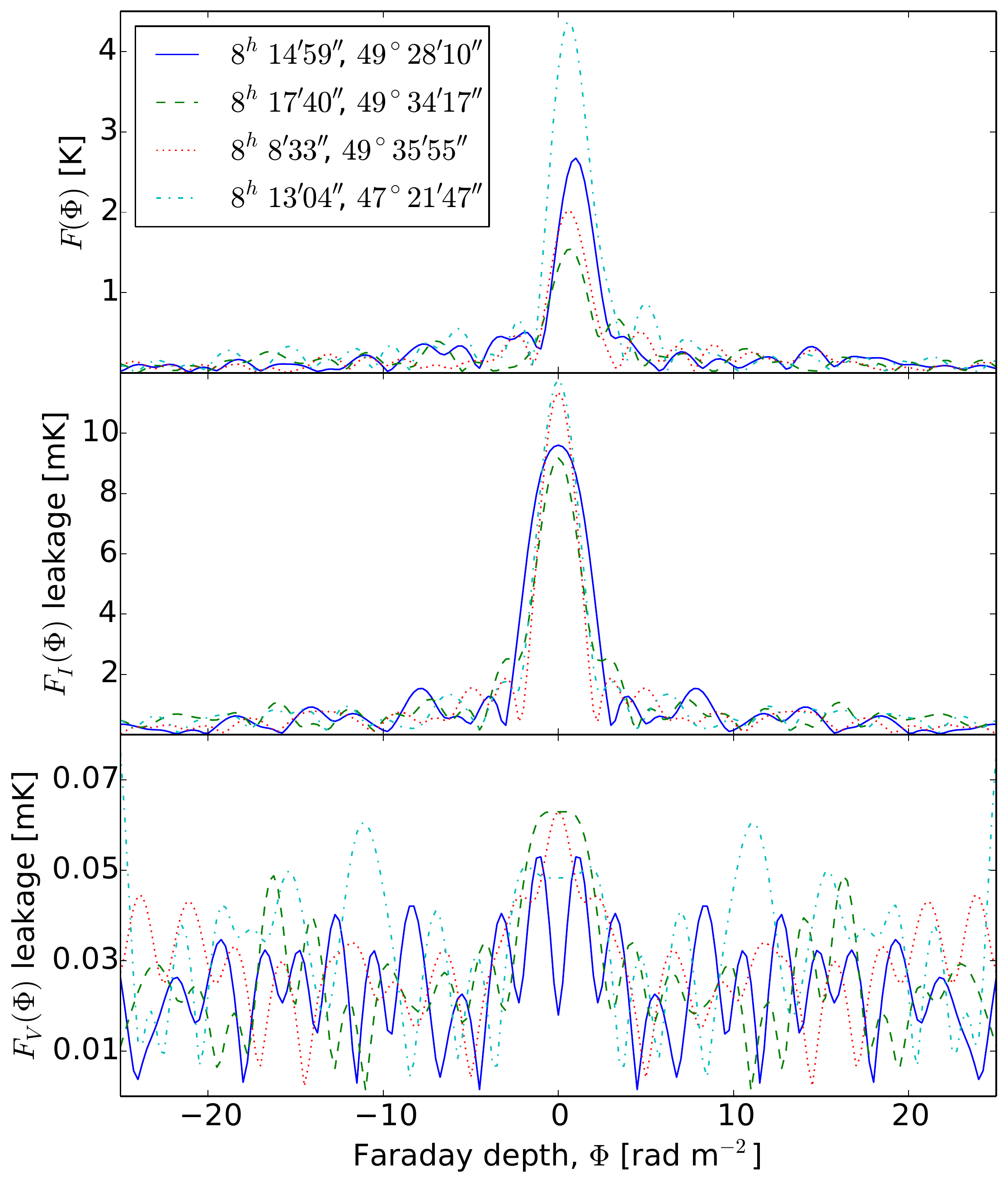}
\caption{Four lines of sight along Faraday depth (Faraday spectrum) for 4 bright pixels in the Faraday dispersion images $F(\Phi)$ (top), leakage into $F_I(\Phi)$ (middle) and leakage into $F_V(\Phi)$ (bottom). The pixels were chosen according to their intensity in $F_I(\Phi)$ and their RA and DEC are shown in the legend.}
\label{f:rmprof}
\end{figure}
The polarization RM-cube, $F(\Phi)$ basically represents the real data that we took as our input model, as leakage from $Q$ to $U$ and vice versa will be very small compared to their brightness.
The maximum intensity in the cube is $\sim 5$ K which is seen at $\Phi=+1.2$ rad/m$^2$ and the brightest structures are located within the Faraday depths of -1.5 and +5.0.
Three slices of the $F(\Phi)$ cube at $\Phi$=0,1,2 rad/m$^2$ are shown in the top 3 panels of Fig. \ref{f:rmimg} and the diffuse Galactic emission is prominent in all of them, but increases toward $\Phi=+1$ rad/m$^2$.
For a detailed analysis of polarized emission in the 3C196 window seen by LOFAR, we refer the reader to Jeli\'{c} et al. (in preparation).
The corresponding slices of $F_I(\Phi)$ leakage are shown in the middle panels of the figure and here we see that the highest leakages appear at $\Phi=0$ and their peak is $\sim 10$ mK.
As differential beams vary slowly with frequency (e.g. see Fig. \ref{f:freq}), the leakages caused by them are a smooth function of frequency, thereby making them localized around $\Phi=0$ in RM space.
This property can be utilized to correct the effects of leakage, but performing a realistic leakage removal is beyond the scope of this paper \citep[e.g. see][]{ge}.
However, in section \ref{s:ddecorr}, we will show the results of a correction that does not take the differential beam into account.
Another aspect of these images can be seen by focusing on the central 2 degrees; for example, although $F(\Phi=0)$ is highest in the central part, the corresponding $F_I(\Phi=0)$ leakage is still much lower than that of the outer region (between the inner and the outer circle).
This is expected as leakage terms of the beam Mueller matrix increase toward the outskirts.

The 3 bottom panels of Fig. \ref{f:rmimg} show the same Faraday dispersion images for leakages into Stokes $V$ and they are much lower than the corresponding leakages into Stokes $I$---so much lower that they are dominated by leaked noise.
This can be understood in terms of the differential beam of Fig. \ref{f:mueller}--- the $M_{42}$ and $M_{43}$ components of this matrix are responsible for leaking $Q,U$ to $V$ and we see that they are 2-3 orders of magnitude lower than the components responsible for leakage into Stokes $I$, i.e. $M_{12}$ and $M_{13}$.

Behaviour of the leakage in Faraday space can be seen more clearly in Fig. \ref{f:rmprof}, where we show 4 lines of sight along Faraday depth (Faraday spectrum) for 4 bright pixels in $F(\Phi)$ (top), $F_I(\Phi)$ (middle) and $F_V(\Phi)$ (bottom).
The bright pixels were chosen in $F_I(\Phi)$ and then the corresponding pixels were found in $F(\Phi)$ and $F_V(\Phi)$.
The fact that instrumental polarization and leakage appears at $\Phi=0$ in a Faraday spectrum, convolved with the RMSF, is evident from the middle panel of the figure.
It is not so evident in $F_V(\Phi)$ due to the dominance of leaked noise.

\subsubsection{3D Power spectra}
\begin{figure*}
\includegraphics[width=0.9\linewidth]{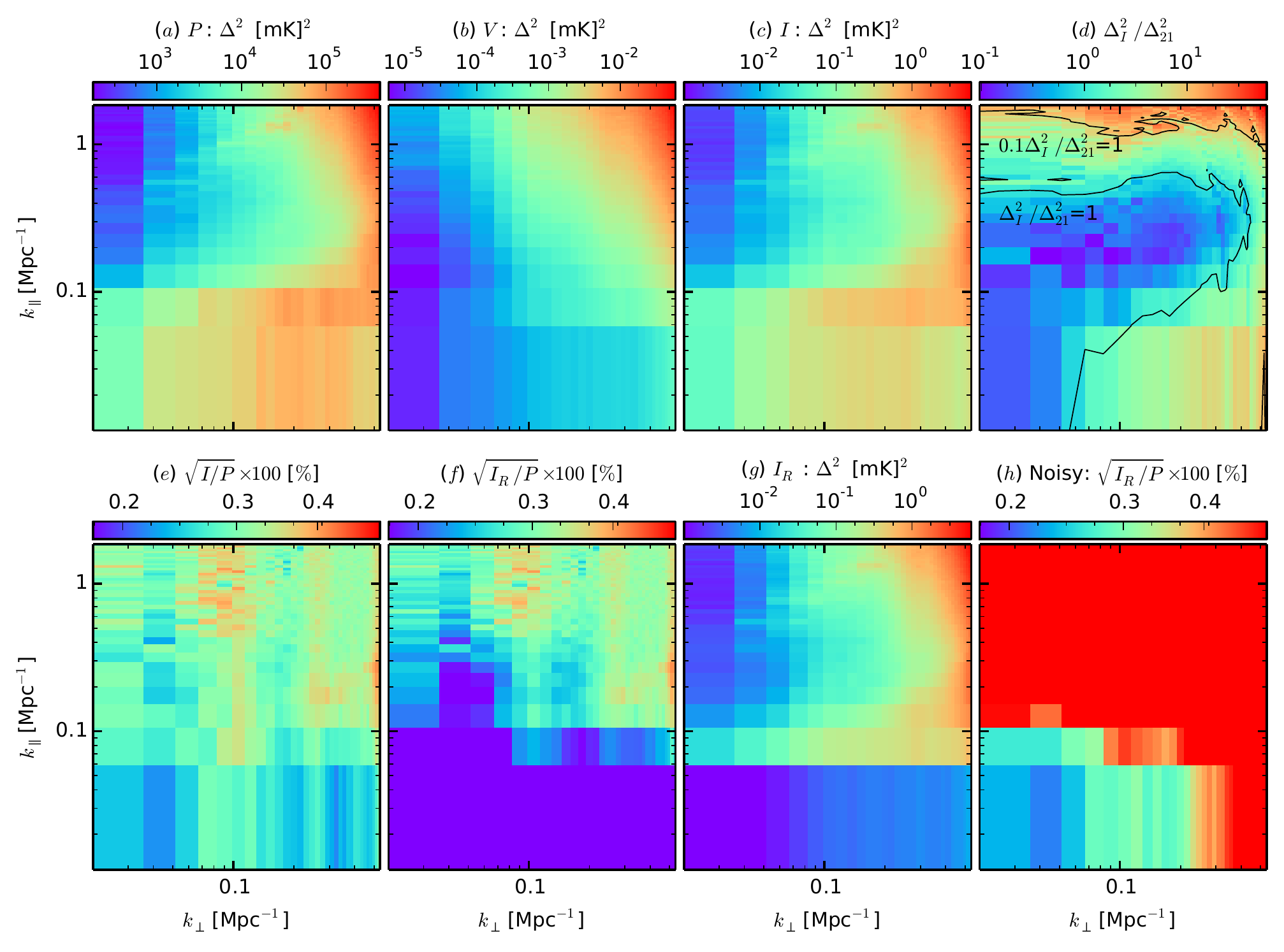}
\caption{\textit{Top}: Cylindrically averaged 3D power spectra (PS) of the polarized emission ($\mathcal{P}$) within the central $4^\circ$ of the 3C196 field ($a$) and its leakages into Stokes $V$ ($b$) and $I$ ($c$).
Panel $d$ shows the ratio between panel $c$ and the corresponding PS of the 21-cm differential brightness temperature $\Delta^2_{21}$ for the fiducial model of \citet{mes} at $z=9$.
The contours are drawn where this ratio is $1$ for both the normal leakage and the leakage reduced by 70\%.
\textit{Bottom}: ($e$) Square-root of the ratio between the panels $c$ and $a$ expressed as a percentage.
The $g$ and $f$ panels represent the same quantity as that of the $c$ and $e$ panels respectively, but for the case when 4 foreground components were removed from the leakage into Stokes $I$.
Panel $h$ shows the same percentage as the $e$ and $f$ panels, but when 60 mK noise was added to the $I$-leakage before foreground removal.
The subscript $R$ stands for GMCA residual.}
\label{f:ps3dc}
\end{figure*}

A 3D power spectrum analysis of the DDE-corrupted image-cubes would be most interesting, as this would allow us to calculate the amount of polarized Galactic foreground leaked into a possible `EoR window' (a region in 3D Fourier space where the EoR signal is taken to be least contaminated) of LOFAR; for an example of an EoR window, see Fig. 1 of \citet{di} that was made using the instrumental parameters of MWA.
In the top panels of Fig. \ref{f:ps3dc}, we show the 3D cylindrical power spectra of the beam-corrupted polarized emission ($\mathcal{P}$), its leakages into Stokes $I$ and $V$, and the ratio between the power spectra of the $I$-leakage and that of the 21-cm differential brightness temperature $\delta T_b$ of the fiducial model of \citet{mes}.
The plots show the power that lies within a given $k_\perp,k_\parallel$ bin in units of [mK]$^2$ or as a ratio.

The $P$ spectrum (panel $a$) exhibits the same characteristics that one would expect based on the behaviour of the polarized emission in RM-space (described in the previous section).
As in RM-space the brightest polarized emission were found near $\Phi=1$ rad/m$^2$, so here the power is high at low $k_\parallel$ ($<0.1$ Mpc$^{-1}$).
Some additional power is seen in a wedge-shaped region at $k_\parallel>0.1$ and $k_\perp>0.1$ which can be attributed to the frequential unsmoothing of the intrinsically smooth polarized foreground by the frequency-varying PSF, and the extra power at high $k_\parallel,k_\perp$ is due to noise.
At $k_\perp<0.04$ power is very low, as expected, and it reaches 
its maximum at around 0.3 Mpc$^{-1}$.
The maximum power is around $4.4\times 10^5$ [mK]$^2$ which is found at the highest $k_\perp,k_\parallel$.

The $I$-leakage spectrum ($c$) looks very similar to the $P$ spectrum, and the leakage power reaches up to $\sim 5.5$ [mK]$^2$.
At $k_\perp<0.1$ and $k_\parallel>0.1$ leakage power is 2--3 orders of magnitude lower than the maximum.
In order to see if $I$ is just a scaled down version of $P$, we calculate the ratio $\sqrt{I/P}$ as a percentage of $P$, shown in panel $e$, which gives an estimate of the percentage of rms leakage at different $k_\perp$ and $k_\parallel$.
Evidently, at $k_\parallel<0.06$ leakage rms is 0.2\%--0.3\% of the polarization rms, and can go as high as 0.4\% at high $k_\parallel$ where noise leaked from $P$ to $I$ dominates.

The $V$-leakage spectrum is shown in panel $b$, and its level is much lower, the peak being around 0.06 [mK]$^2$.
The region at high $k_\parallel$ and high $k_\perp$ is dominated by noise, as signal-to-noise ratio is lower for longer baselines.
By comparing this spectrum with that of $P$, it can be seen that the rms of the $V$-leakage at low $k_\parallel$ is only $\sim 0.003\%$ of the rms of the polarized emission which means that the uncorrected $V$ leakage is negligible compared to the current noise levels in the EoR experiments within a FoV of $4^\circ$.

As noted in the previous section, leakage is lower near the centre of the field (see Fig. \ref{f:rmimg}). In order to quantify the associated decrease in power, we calculated the power spectra of $I$, $P$ and $\sqrt{I/P}$ within the inner $3^\circ$ of the field. We found that, in this case, maximum leakage into $I$ at low $k_\parallel$ is $\sim 4.9$ [mK]$^2$ and $\sqrt{I/P} \approx 0.2\%$ at $k_\parallel < 0.06$ which is lower than the level of leakage within the inner $4^\circ$.

\begin{figure}
\includegraphics[width=0.9\linewidth]{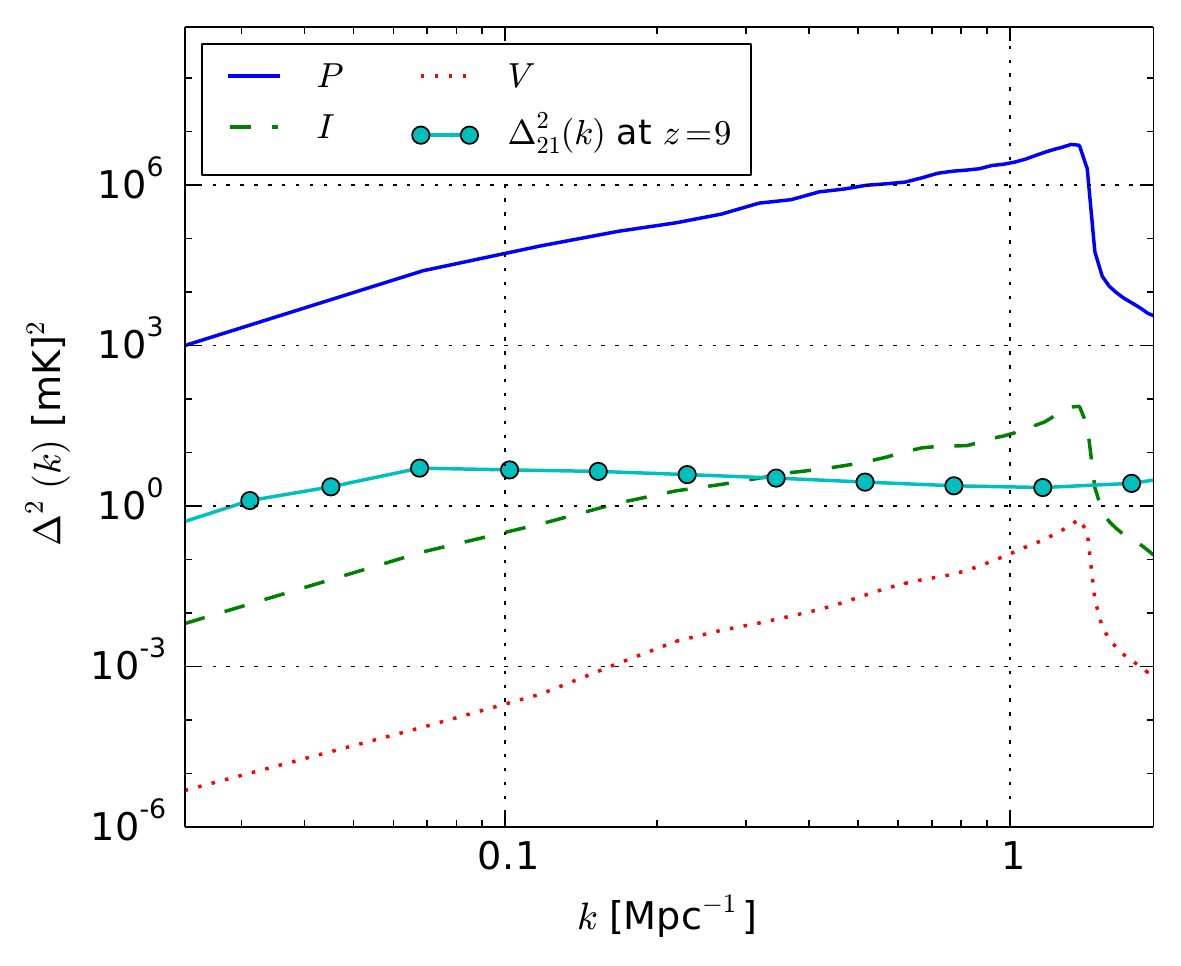}
\caption{Spherically averaged 3D power spectra of the polarized emission within the central $4^\circ$ of the 3C196 field (top solid), its leakages into Stokes $I$ (middle solid) and $V$ (bottom solid) caused by the LOFAR model beam, and the PS of the 21-cm differential brightness temperature $\Delta^2_{21}$ at $z=9$ (solid with circles) for the fiducial model of \citet{mes}.}
\label{f:ps3ds}
\end{figure}

To see how $P\rightarrow I$ leakage affects the EoR signal, we took the 3D spherical power spectrum of the fiducial model of 21-cm differential brightness temperature $\delta T_b$ at $z=9$ from \citet{mes} and calculated the corresponding cylindrical power spectrum as $\Delta^2_{21}(k_\perp,k_\parallel)=k_\perp^2 k_\parallel \Delta^2_{21}(k)/[2(k_\perp^2+k_\parallel^2)^{3/2}]$.
The spherical PS is plotted in Fig. \ref{f:ps3ds} along with the spectra of $P$, $I$ and $V$, and the figure clearly shows that the EoR signal power is higher than the $I$-leakage at $k<0.3$ Mpc$^{-1}$, and can be 2 orders of magnitude higher at the lowest scales.
The ratio between $\Delta^2_{21}(k_\perp,k_\parallel)$ and $\Delta^2_I(k_\perp,k_\parallel)$ is plotted in Fig. \ref{f:ps3dc}d where the contours are drawn at $\Delta^2_I/\Delta^2_{21}=1$ for the normal case and the case when 70\% of the leakage had been removed.
Evidently, there is an `EoR window' above the PSF-induced wedge and below $k_\parallel\sim 0.5$ Mpc $^{-1}$, and the window extends up to $k_\parallel\sim 1$ Mpc$^{-1}$ when 70\% leakage is removed.

\subsection{Polarization leakage removal} \label{s:ddecorr}
As \citet[][section 7.2]{je10} have discussed at length, a leakage of the polarized foreground into total intensity will be a major obstacle in detecting the EoR signal if (1) the level of leakage is comparable to the intensity of the EoR signal, and/or (2) frequency spectrum of the leakage mimics that of the signal.
Fortunately, in the 3C196 field the latter is not the case, as we have seen that there is no significant polarization at high Faraday depths, or, equivalently, at high $k_\parallel$.
However, the power of leakage could be comparable to that of the signal at high $k_\parallel$ and hence leakage needs to be removed with sufficient accuracy to extend the EoR window.

There are many methods for removing foregrounds from Stokes $I$; some assume spectral smoothness of the foreground and try to fit it out using polynomials, while others do not assume anything and hence are called `blind' or non-parametric methods \citep[for a list see][]{ch14}.
The best way to remove the leakage contribution of the foreground is, of course, to use the time-frequency-baseline dependent Mueller matrices during calibration and/or imaging to produce beam- and leakage-corrected images.
Another potential way is to correct them in the Faraday dispersion images, i.e. correcting the $F_I(\Phi)$ using information from $F(\Phi)$ as demonstrated by \citet[][see section 6.2 and Fig. 6]{ge}.
For leakages as smooth as in the field of 3C196, simply filtering the $F_I(\Phi)$ for $\Phi\sim 0$ could be another potential solution.
However, testing these methods is beyond the scope of this paper, and here we use a non-parametric foreground removal method, called GMCA (generalized morphological component analysis; \citealt{bo07,bo08a,bo08b}), that has been shown to be able to remove foregrounds from simulated LOFAR-EoR data with high accuracy \citep{ch13}.

If a signal is represented as $X=AS+N$ where $S$ is the foreground to be extracted, $N$ is noise and $A$ is the mixing matrix, then GMCA tries to calculate a mixing matrix for which $S$ is sparsest (have the least number of non-zero wavelet coefficients) in the wavelet domain.
For details of the algorithm we refer the readers to \citet{ch13}.
We run GMCA on the Stokes $I$-leakage cube to extract and subtract 4 components of the leaked foreground, as this number has been shown to yield good results \citep{ch14}, and produce 3D cylindrical power spectrum from the residual cube which is shown in Fig. \ref{f:ps3dc}g.
It clearly shows that the power of the smooth foregrounds at low $k_\parallel$ ($<0.06$) has been reduced by almost two orders of magnitude by GMCA; compare it with the input $I$-leakage spectra of panel $c$ that is plotted on the same scale.
On the other hand, everything above $k_\parallel=0.1$ has been kept completely untouched due to low SNR---where S is the foreground and N is the noise including the cosmic signal---as GMCA cannot produce reliable model for the foregrounds when the SNR is low.
In panel $f$, we plot the ratio of the GMCA residual PS and the polarization PS which shows that after GMCA subtraction, rms residual leakage at $k_\parallel<0.1$ is around 0.1\% of the polarized intensity.
However, the EoR signal could also be removed along with the foreground in this case as there was no noise in Stokes $I$ except for a very low level of noise leaked from $Q,U$.

To see how additive noise affects the removal of leakage, we add 60 mK (rms) noise, which should be reached after 600 hours of integration using LOFAR \citep{ch14}, to the Stokes $I$ leakage maps at all frequencies.
The noise was added to the visibilities and a new image cube was produced from the noisy visibilities.
We run GMCA on the noisy $I$-leakage cube and produce a 3D cylindrical PS from the residual and take the square-root of the ratio of this PS with respect to the $I$-leakage which is shown as a percentage in Fig. \ref{f:ps3dc}$h$.
We see that almost no leakage has been removed in this case, not even in the relatively high SNR region at low $k_\perp$.
Therefore, we conclude that in case of such levels of noise, either a different strategy should be taken to remove foreground-leakage, or the leakage dominated region (where the leakage is more than the EoR signal) should be avoided to some extent (see \citet{ch14} for a discussion on the relative merits of foreground removal and avoidance.)

\section{Summary and conclusions} \label{s:summ}
This paper presents a first step in the analysis of the systematic errors of a radio interferometer, with a focus on polarization leakage, by simulating the LOFAR observations of both compact and diffuse emission in the presence of direction independent and direction dependent errors which are treated separately.
We have revisited the measurement equation of a radio interferometer and modelled the direction independent (DI) and direction dependent (DD) errors as $2\times 2$ Jones matrices and the corresponding $4\times 4$ Mueller matrices have been used to show the polarization properties of the instrument.
The full polarization DD-Mueller matrix (Fig. \ref{f:mueller}) describing the time-frequency-direction dependent behaviour (e.g. see Fig. \ref{f:mueller} and \ref{f:freq}) of the response of a baseline of LOFAR, created by two stations (Fig. \ref{f:station}), has been presented to be a DD equivalent of the DI-Mueller matrix of Eq. \ref{eq:dieM}.

We have simulated an observation with DI-errors by assuming them to be random at every timestep and the rms of the random numbers drawn from a Gaussian distribution with zero mean is dubbed the `rms DI-error'.
We find that self-calibration can solve for these errors to an extremely high accuracy if the sky model is perfect as, in that case, the information provided by an interferometer will be highly redundant.
For an rms DI-error of $10^{-3}$, the selfcal error is less than 0.002\% and the corresponding error in the rms of the resulting residual image is less than 0.005\% (Fig. \ref{f:die}).

The only DD-error that we have simulated is the differential (normalized with respect to the phase centre) station beam of LOFAR.
We simulated the LOFAR observations of extragalactic unpolarized point sources in the 3C196 observing window of the LOFAR-EoR experiment including the DD-errors and estimated the flux and position errors due to self-calibration with incomplete sky models and the percentage of $I\rightarrow (Q,U)$ leakage of the brightest sources (see Fig. \ref{f:3c}).
We see that the errors go down significantly as the sky model is improved.
However, calibrating with only unpolarized sources has its limitations, e.g. the unitary ambiguity \citep{wi,ca14}.
There is no plan for using polarized sources in calibrating LOFAR EoR data until now, as there are very few intrinsically polarized point sources in the data, and the polarized emission is dominated by diffuse emission \citep[e.g. see][]{ya13,je14}.
We test a possible strategy of correcting the DD errors from point sources using {\tt AWImager} with an unrealistic, exaggerated sky model and see that {\tt AWImager} can remove up to 80\% of the leakage from Stokes $Q,U$, but a more elaborate testing of this algorithm with realistic sky models has to be done to reach any final conclusion.

To predict the level of polarization leakage in the Stokes $I$ images of the 3C196 field, we took the real LOFAR observations of Galactic diffuse polarized emission in this field and created an unreal sky model where $I=V=0$ to quantify the leakages from $Q,U$ to $I,V$ caused by the DD errors.
An RM-synthesis of the DDE-corrupted $\mathcal{P}$ image cubes showed that in this particular field polarization peaks within the Faraday depths ($\Phi$) of -1 and +5 rad/m$^2$.
From the effective Stokes $I$ Faraday dispersion images we saw that polarization leakage is localized around $\Phi=0$ (Fig. \ref{f:rmprof}), as DD-errors do not have any rapid variation along frequency.
Maximum leakage was found to be around 15 mK which could be comparable to the EoR signal (Fig. \ref{f:rmimg}).

To understand the level of leakage contaminant in the `EoR window' of the instrumental $k$-space, we calculated the cylindrically and the spherically averaged 3D power spectra (PS) of $I,P,V$ cubes.
The $P$ spectrum shows characteristic smooth polarized foregrounds at low $k_\parallel$ (Fig. \ref{f:ps3dc}) and the $I$-leakage spectrum looks very similar to this.
From the power ratio, $\sqrt{I/P}$ we showed that the percentage of rms leakage over the $k_\perp,k_\parallel$ space varies by a factor of 2 and ranges from 0.2\% to 0.4\%.
We compared the $I$-leakage with the 3D PS of the expected 21-cm differential brightness temperature at $z=9$ simulated by \citet{mes} and saw that the region above the PSF-induced wedge and below $k_\parallel\sim 0.5$ Mpc$^{-1}$ is dominated by the cosmic signal (Fig. \ref{f:ps3dc}d) and hence defines a potential `EoR window', and the window expands substantially after removing 70\% of the leakage.

As the $I$-leakage do not mimic the EoR signal in this case, we tried to remove it using GMCA which is being used to remove diffuse foreground from the LOFAR-EoR data.
From the 3D PS of the residual left after the removal of foreground leakage components by GMCA, we saw that (Fig. \ref{f:ps3dc}f,g) at $k_\parallel<0.1$, i.e. in the high SNR regime, GMCA could reduce the leakage by up to two orders of magnitude while the region above that scale was left completely untouched.
For a more realistic analysis, we added 60 mK noise to the Stokes $I$ leakage maps, reran GMCA on it and saw that (Fig. \ref{f:ps3dc}h) in this case almost no leakage was removed, not even in the relatively high SNR region.

Antennas for the future arrays like SKA, that have EoR detection as one of the main scientific objectives, are being designed in such a way that their polarimetric performance is good enough to be able to minimize the effects of polarization leakage (de Lera Acedo; private communication).
A recently proposed figure of merit for quantifying the polarimetric performance is the \textit{intrinsic cross-polarization ratio} (IXR) which, in Mueller formalism, can be directly related to the instrumental polarization \citep[][eq. 23]{ca11}.
Our LOFAR results show an instrumental polarization of around 0.3\% (Fig. \ref{f:ps3dc}e; ignoring $V\rightarrow I$ leakage) within the FWHM of the nominal station beams, i.e. within a FoV of $\sim 4^\circ$.
This corresponds to an IXR$_\text{M}$ (\textit{Mueller IXR}) of 25 dB, or equivalently an IXR$_\text{J}$ (\textit{Jones IXR}; see eq. 25 of \citealt{ca11}) of 56 dB, and if the leakage can be reduced by 70\%, IXR$_\text{M}$ will improve to 35 dB.
Therefore, we can say that if SKA has a minimum IXR$_\text{M}$ of 25 dB within the central $\sim 4 ^\circ$ of its nominal station beams, then even a modest polarimetric calibration ($\sim 70\%$ leakage removal) will ensure that the polarization leakage remains well below the expected EoR signal at the scales of 0.02--1 Mpc$^{-1}$.
However, if the IXR$_\text{M}$ is lower within a FoV of $4^\circ$, more leakage needs to be removed to reach the same level as before in relation to the EoR signal in the power spectra, e.g. if the IXR$_\text{M}$ is 20 dB, 91\% leakage has to be removed, and if it is 15 dB, 97\% has to be removed.

The major conclusions of this paper are the following.
\begin{enumerate}
\item Two properties of the polarization leakage can be utilized for its removal in this specific case: it appears around a Faraday depth of 0 rad/m$^2$ in RM-space and the overall variation of the rms of the fractional leakage in the instrumental $k$-space is less than a factor of 2.
\item In the cylindrically averaged 3D power spectra, a clear `EoR window' can be defined in terms of polarization leakage above the wedge and below $k_\parallel\sim 0.5$ Mpc$^{-1}$. Within this window, the EoR signal dominates the polarization leakage and the window takes up the whole $k$-space at $k_\parallel<1$ after removing 70\% of the leakage.
\item A DDE-blind foreground removal method like GMCA is not ideal for removing leakage of diffuse polarized emission, as the level of leakage is lower than the current noise level in the LOFAR observations.
\end{enumerate}

\section*{Acknowledgments}
We thank the anonymous reviewer for his/her useful comments.
KMBA, LVEK, AG and HKV acknowledge the financial support from the European Research Council under ERC-Starting Grant FIRSTLIGHT -- 258942.
VJ acknowledges the financial support from The Netherlands Organization for Scientific Research (NWO) under VENI grant -- 639.041.336.
AGdB, SBY and VNP acknowledge support from the European Research Council under grant 399743 (LOFARCORE).
AHP and SZ acknowledge the support from Lady Davis Foundation and NWO VICI grant.
GH acknowledges funding from the the People Programme (Marie Curie Actions) of the European Union's Seventh Framework Programme (FP7/2007-2013) under REA grant agreement no. 327999.
ITI was supported by the Science and Technology Facilities Council [grant numbers ST/F002858/1, ST/I000976/1 and ST/L000652/1].
In addition, KMBA would like to thank George Heald and Cyril Tasse for useful suggestions.

\footnotesize{
\bibliographystyle{mn2e}

}

\label{lastpage}
\end{document}